\documentclass[acmsmall,screen,review=false]{acmart}

\usepackage{hyperref}
\usepackage{graphicx}
\usepackage[ruled]{algorithm2e}
\usepackage{subfigure}
\usepackage{multirow}
\usepackage{caption}
\usepackage{accents}

\newcommand{\name}{DecVi} 
\newcommand{\namens}{DecVi}

\AtBeginDocument{%
  }

\setcopyright{acmcopyright}
\copyrightyear{2022}
\acmYear{2022}
\acmDOI{XXXXXXX.XXXXXXX}

\acmJournal{JACM}
\acmVolume{37}
\acmNumber{4}
\acmArticle{111}
\acmMonth{8}

\begin{document}

\title{\name: Adaptive Video Conferencing on Open Peer-to-Peer Networks}

\author{Jingren Wei}
\authornote{Both authors contributed equally to this research.}
\email{wei.1276@osu.edu}
\author{Shaileshh Bojja Venkatakrishnan}
\authornotemark[1]
\email{bojjavenkatakrishnan.2@osu.edu}
\affiliation{%
  \institution{The Ohio State University}
  \city{Columbus}
  \state{Ohio}
  \country{USA}
  \postcode{43210}
}

\begin{abstract}
Video conferencing has become the preferred way of interacting virtually, especially since the pandemic.
Current video conferencing applications---the likes of Zoom, Teams, Webex or Meet---are centralized, cloud-based platforms whose performance crucially depends on the proximity of clients to their data centers. 
Clients from low-income countries are particularly affected as most data centers from major cloud providers are located in economically advanced nations (e.g,. Google has no data centers in Africa).
Centralized conferencing applications also suffer from occasional massive outages and are embattled by  serious privacy violation allegations. 
In recent years, decentralized video conferencing applications built over peer-to-peer networks and incentivized through blockchain technology are becoming popular. 
A key characteristic of these networks is their openness: anyone can host a media server on the network and gain monetary reward (through crypto-tokens) for providing conferencing service.  
Strong economic incentives combined with a relative low entry barrier to join the network, makes increasing server coverage to even remote regions of the world possible with these applications. 
On the other hand, these very same reasons also lead to a security problem---a media server may obfuscate its true location (e.g., using proxy servers or VPNs) in order to gain an unfair business advantage. 
In this paper, we consider the problem of multicast tree construction for video conferencing sessions in open, p2p conferencing applications. 
We propose \name, a decentralized multicast tree construction protocol that adaptively discovers efficient tree structures based on an exploration-exploitation framework. 
\name\ is motivated by the combinatorial multi-armed bandit problem and uses a succinct learning model to compute effective actions. 
Despite operating in a multi-agent setting with each server having only limited knowledge of the global network and without cooperation among servers, experimentally we show \name\ achieves similar quality-of-experience compared to a centralized globally optimal algorithm while achieving higher reliability and flexibility.

\end{abstract}




\maketitle

\section{Introduction}

Video conferencing has become an ubiquitous method of virtual interaction for many--with millions of meetings happening daily for work, education and leisure--especially since the global pandemic. 
The global video conferencing market was valued at more than 6 billion US dollars in 2021, and is projected to rise to 14 billion US dollars by 2029~\cite{fortunebusiness}. 
Emerging technologies notably the Internet-of-Things and augmented/virtual reality are further expected to drive this growth over the coming years~\cite{vrvideoconffuture, iotvidconference, chakraborti2018using}. 

A bulk of the video conferencing sessions today occurs over cloud-based platforms such as Zoom, Microsoft Teams, Google Meet and Cisco Webex. 
In these systems, media servers housed in massive data centers act as central points of aggregation for receiving and distributing video streams across participants of a conferencing session~\cite{chang2021can}. 
Fueled by the emergence of blockchain technology, 
 in recent years a number of decentralized and open peer-to-peer (p2p) video conferencing platforms are also being developed in the industry~\cite{huddle01,livepeer,crewdle,impervious,jami,gem4me,p2pchat,toxchat,utopia,Peer5,keetio,thetanetwork}. 
In an open p2p conferencing system, media servers owned by individuals serve clients at various geographical locations by relaying their streams to appropriate endpoints in exchange for a fee provided through a blockchain.
E.g., Livepeer's decentralized network includes servers from individuals with their aggregate capacity exceeding 70,000 GPUs  today~\cite{livepeer}.  
While still in their infancy~\cite{huddle01,livepeerstate}, decentralized conferencing solutions carry unique benefits making them promising as alternative platforms to existing centralized counterparts. 
Some of the important benefits include: 
\begin{itemize}
\item {\em Quality-of-experience (QoE).}  
Centralized video conferencing frameworks employ only a limited number of data center regions (fewer than 20, worldwide~\cite{webexlocations}) for media service, which causes disruptions to sessions under high demands, and excessive streaming lag.   
E.g., a recent study~\cite{choi2022zoom} reports Zoom users in Canada experiencing significant disruptions under a surge in demand, due to Zoom's media servers being overloaded.   
Also, the extent of lag clients experience exhibits considerable variation depending on where clients are located. Another study~\cite{chang2021can} shows clients on the west coast in the US experiencing more delay compared to clients on the east coast on Zoom and Webex, as only media servers on the east coast are used for all conferences in these platforms (in their free-tier service category).  
The latency between two users both located on the west coast is therefore higher than the latency between a west coast user and an east coast user. 
It is also reported that Zoom and Webex predominantly use data centers in the US even for routing international traffic, resulting in European clients seeing 55–75 ms and 45–65 ms higher
median lags on Zoom and Webex respectively. 
Further, excessive jitter due to long packet travel times affects the fluidity of conversations, and causes ``Zoom fatigue"~\cite{boland2021zoom,fauville2021zoom}.  
Low-income nations distant from major data center regions suffer even more. 

In p2p video conferencing systems, on the other hand, 
the relatively low entry barrier for hosting servers creates incentives for placing servers at diverse geographical locations and close to client hot spots, even in developing nations~\cite{jitsimeetself}.
Increased proximity of servers to clients leads to lower packet latencies, lesser jitter and fewer frames dropped, compared to the status quo thus improving QoE.  
The turnaround time for augmenting the network with more server capacity is also faster in a p2p network, compared to centralized systems. 

\item {\em Availability.}
In a p2p network it is unlikely for multiple, independent servers to go down at the same time, minimizing chances of a network-wide outage. 
Whereas in centralized platforms outages are not uncommon, affecting millions of clients each time~\cite{zoomdown,teamsoutage,meetdown}. 

\item {\em Pricing.} A p2p conferencing system is also likely to be cheaper (by up to 10--100$\times$ according to reports~\cite{livepeerpricest}) for end users due to a lack of need for a dedicated server infrastructure  and a more open, transparent market.
Centralized conferencing systems, on the other hand, require adequate server provisioning that must be paid for even if the servers are not being fully used; this cost ultimately get passed on to end users~\cite{zoomcostmaint,zoomplansandprice}. 

\item {\em Privacy and censorship.}
There is also a growing concern about the privacy and censorship practices of centralized video conferencing providers~\cite{zoombanspace,zoombgpr}.  
In contrast, the open nature of p2p conferencing systems makes it a fully transparent medium where no single party has the ability to unilaterally censor users or collect their data.  
\end{itemize}

A key challenge in building a large scale and open p2p conferencing network is how to determine efficient paths over which to deliver client media streams. 
For good QoE, a streaming path must have low latency and sufficient bandwidth available to carry video chunks of appropriate quality, as requested by a receiving client's application.   
Further, servers often have only limited compute and bandwidth capacities which restrict the maximum number of streams that can be relayed by any one server.  
Thus, for conferencing sessions involving a large number of clients it becomes necessary to route streams over multiple (potentially geographically separated) servers structured as a multicast tree.
Determining a good subset of servers for the multicast tree, and how to structure the multicast tree to maximize QoE, are nontrivial problems whose solution depends on various factors including the geographic location of clients and servers, video quality desired by each client, network bandwidth available at the clients and servers, and amount of compute available at the servers.  
These factors are not only highly heterogeneous across peers but can also be time varying which further complicates the problem. 

In an open system, clients and servers are free to join or leave the network at will. 
The potentially large number of servers and peer churn make it impractical for a central entity to keep track of the servers and compute the routing paths for clients. 
Centralized route computation is also a single point of failure which diminishes system availability under an attack. 
It is therefore desirable to use a fully decentralized route computation mechanism.

We present \namens, a fully decentralized and efficient route computation algorithm for large scale, p2p conferencing networks. 
In \name\ a multicast tree of servers is built for each source of a video stream in a conferencing session, with the source as the root and all other clients in the session as leaves.
A multicast tree is initialized arbitrarily, but is iteratively refined over time using an exploration-exploitation framework for increasing client QoE.   
\name\ is inspired by the combinatorial multi-armed bandit problem~\cite{chen2013combinatorial,slivkins2019introduction}, with a server in a multicast tree treating downstream servers to whom it must forward streams to and the corresponding stream qualities, as the `arms' of a bandit problem. 
The streaming latency and quality of stream received by a client form the `reward' earned, which is communicated by the client to the multicast tree servers as feedback. 
The adaptive approach of \name\ results in multicast trees that are automatically tuned to the various heterogeneities in the network without explicit human input. 
Our proposed protocol can scale to networks of several thousand clients and servers due to its decentralized design. 

Prior works have proposed distributed algorithms based on analytical models for multiparty video conferencing with multiple media servers for relaying~\cite{liang2011optimal,zhao2013enabling,ponec2009multi}. 
A model-based algorithm requires clients and servers to have knowledge of the model parameters (available bandwidth, number of servers etc.) to run.
Moreover, the model itself may not be a good approximation of the system. 
In contrast, \name\ is an entirely adaptive solution which does not require clients or servers to have knowledge of the system parameters. 

More recently, a number of works have presented p2p conferencing designs for WebRTC~\cite{ng2014p2p,nurminen2013p2p}. 
These designs, however, consider only a single media server for connecting the clients. 
With the development of scalable video coding~\cite{bakar2018motion}, selective forwarding units (SFUs) are emerging as lightweight, low-delay media servers (with forwarding delay typically $<$ 20ms~\cite{amirante2015performance,andre2018comparative,baltas2014ultra}). 
The low complexity of SFUs has encouraged operators to deploy a cascade of SFUs for geo-distributed conferencing sessions for efficiency~\cite{grozev2019efficient,vidyoiocascade}. 
Dynamically deciding which SFUs to utilize for a  conferencing session with cascaded SFUs is still an active area of research. 
Our work addresses this problem for open, permissionless systems where anyone is free to join the network and host SFUs. 

The contributions of this paper are: 
\begin{enumerate}
\item We formulate multicast tree construction for video conferencing in open, p2p systems as a distributed learning problem. 
A learning approach to p2p design for streaming video  has not been previously proposed, to our best knowledge. 
\item We propose a decentralized algorithm for tree construction inspired by the combinatorial multi-armed bandit problem.
However, our work is different from standard multi-agent bandit settings which consider all agents working on instances of the same bandit problem or sharing the same reward function globally.
The algorithm and empirical results we provide may be of independent interest to the multi-agent bandit research community. 
\item Experimentally, we build a custom event-driven simulator and show that \name\ achieves a 19\% latency speedup compared to centralized conferencing applications whose multicast trees are structured as a star topology with a data center as the hub node. 
\item In smaller conference settings (5 to 10 clients involved), \name\ achieves the globally optimal topology, while in larger settings (with 50 clients) it achieves about 85.2\% of QoE performance compared to an omniscient centralized baseline. 
\end{enumerate}

\section{Background}

\subsection{Video Stream Compression and Encoding} 
Motivated by emerging p2p video conferencing systems~\cite{huddle01,livepeer,thetanetwork}, we consider a system where clients in a conferencing session are connected to an open network of media servers for sending and receiving media streams (Figure~\ref{fig:background}).
Servers are required as directly sending streams from each client to all other clients consumes too much bandwidth at the clients for large conferences. 
Even with servers, transporting raw video over the Internet is  bandwidth intensive---it is common for 
conferencing systems today to use a video compression protocol (e.g., H.264/AVC, VP9, H.265/HEVC and AV1 are popular video compression standards) for reducing video bandwidth by 1000$\times$ or more.
Compression must be done at a bit rate that suits the video resolution and quality desired by receiving clients. 
If clients have heterogeneous bit rate requirements, then it becomes necessary to simultaneously encode video at different bit rates, sending an appropriate bit rate stream to each client. 
Traditionally, the computationally expensive operation of  encoding video in to multiple bit rates (called transcoding) was done at a media server, which increased the complexity and cost of media servers.  
A key computational challenge here is that the transcoding operation must be repeated for each distinct bit rate desired.  
Scalable video coding is therefore becoming a popular alternative for achieving heterogeneous bit rate requirements with a low media server complexity~\cite{schwarz2007overview,liu2020grad}. 
In scalable video coding, video is encoded in to multiple bit stream ``layers'', including a base layer and a sequence of enhancement layers. 
Video can be decoded using the base layer and any number of consecutive enhancement layers, with more enhancement layers providing a greater video quality. 
Compared to single-layer coding schemes, scalable video coding incurs a small bit rate overhead (about $\sim$10\%) for the same video quality; 
it has been included as extensions to the popular H.264/AVC, H.265/HEVC, VP9 and AV1 codecs and is beginning to see widespread adoption in mainstream video conferencing applications~\cite{zoomscale,googlevp9}.  
With scalable video coding, media servers are significantly reduced in complexity performing only the job of forwarding video packets without any transcoding. 
Media servers are also called as selective forwarding units (SFU) in a scalable video coding system---depending on the video quality desired by a client an SFU chooses the base layer and an appropriate subset of enhancement layers and forwards to the client. 

\subsection{{Open P2P Video Conferencing}}
Decentralized video conferencing systems with multiple geo-distributed servers are naturally more scalable and have lower streaming delays compared to centralized systems.  
While decentralized conferencing systems have been deployed in the past, they are typically closed systems managed by a single central authority~\cite{trueconf,matrix}.
The economic models of these systems make a pervasive deployment of media servers prohibitively expensive, limiting client QoE in many regions. 
With the development of blockchains and smart-contract technology, decentralized video conferencing systems not managed by any one entity are emerging.
These systems are characterized by their: (1) {\em openness}---any operator may freely join the system and use their servers for providing conferencing service to clients, (2) {\em incentives}---operators earn monetary compensation (typically through crypto-tokens) for providing computation and bandwidth services to clients, and (3) {\em security}---parties cannot deviate from protocol to obtain unfair monetary gains. 
Verification that a server has correctly provided service for a conference occurs through cryptographic primitives on smart contracts. 
These smart contracts are either directly implemented over a blockchain~\cite{livepeer} or implemented off-chain~\cite{keetio,poon2016bitcoin} for additional scalability. 

Clients and media servers discover IP addresses of other media servers through a peer discovery mechanism using gossip. 
Server IPs can also be published on chain as part of the smart contract between clients and servers.
An open system is susceptible to Sybil attacks where an adversary spawns a large number of SFU process instances (assigning fake identities to each instance) over potentially multiple server machines~\cite{swathi2019preventing}.
E.g., using proxies/VPN a server may claim to be at a location which is different from its true location to gain unfair rewards. 
Whereas selecting media servers in close proximity to clients is crucial for achieving low streaming delay. 
Existing open p2p systems use a reputation mechanism realized through delegated staking (i.e., participants 
 stake their personal funds vouching  for the correct operation of certain servers based on past experience with the servers) to identify reliable servers with high available bandwidth.
Even so, determining a media server's true location based on a client's past experience with the server, is considerably challenging and has not been implemented. 
\namens's adaptive server selection protocol is therefore well suited to open p2p settings. 

\section{System Model}
\label{s:model}
We consider a video conferencing session between a set of clients $C$, streamed using a subset of media servers from a set $S$ of available media servers. 
Each client $c\in C$ is the source of a 
 video stream that it seeks to distribute to all the other clients in $C$. 
A client distributes its stream by forming a multicast tree with the client as the root, all the other clients as the leaves, and one or more media servers as the interior nodes of the tree (Figure~\ref{fig:backgroundandmotivatation}). 
Each client forms its multicast tree independently of the multicast trees of other clients.\footnote{Our approach can be generalized to construct a single, common multicast tree to disseminate all streams. We leave a systematic study of such an approach to future work.} 
In the remainder we therefore restrict ourselves to a single client source $c_0 \in C$, and focus on how to construct an efficient multicast tree for $c_0$.  

\subsection{Network Model}
We model the streaming network as a graph $G$ with SFUs $S$ and clients $C$ forming the nodes. 
For two nodes $u, v \in G$, there is a link $(u,v)\in G$ between the nodes if $u$ sends a stream to $v$ in the multicast tree for $c_0$.
Each link $(u,v) \in G$ has a latency $l(u,v) \geq 0$, which is the time it takes for packets sent from $u$ to reach $v$. 
Node pairs $u', v'$ that do not form a link, i.e., $(u', v') \notin G$, also have an associated latency $l(u', v')$ which is the time it takes for packets to go from $u'$ to $v'$ if $u'$ were to forward the stream to $v'$. 
We do not model jitter or packet loss in this model.
Each SFU $s\in S$ has a bandwidth limit of $b_s \in \mathbb{N}=\{0, 1, 2, \ldots\} \geq 0$, which is the maximum rate at which it can upload streams to other nodes. 
We assume download bandwidth is not  constrained, as download bandwidth is typically greater than upload bandwidth in practice.  
Nodes a priori do not have any knowledge of the link latencies or bandwidth limits of other nodes in the network. 
However, we assume the IP addresses of SFUs $S$ is public knowledge and an SFU can connect with any other SFU. 

Video is encoded through scalable video coding at the source with a maximum of $Q \in \mathbb{N}$ layers. 
We assume each layer (base or  enhancement) in the stream consumes 1 unit of bandwidth while uploading or downloading. 
An SFU $s \in S$ that receives $q$ layers of the stream may forward at most $q$ layers to each of its downstream nodes in the multicast tree. 
Moreover, the total number of layers forwarded cannot exceed the bandwidth $b_s$ available at the server. 
Each client $c\neq c_0, c\in C$ has a requirement for receiving at least $q^*(c) > 0, q^*(c) \leq Q, q_c \in \mathbb{N}$ layers, with a packet latency as small as possible. 
Latency of a path $c_0, s_0, s_1, \ldots, s_k, c$ in the multicast tree is the overall delay $l(c_0, s_0) + \sum_{i=0}^{k-1} l(s_i, s_{i+1}) + l(s_k, c)$ incurred by video packets from the source to the destination client. 
Note that each client has a unique path from $c_0$ in the multicast tree. 
For simplicity we do not consider the forwarding delay incurred at each SFU for buffering, processing and relaying video frames; nor do we consider packet transmission delays during upload or download at a node. 
These details can be included in the model without causing any significant changes to the proposed algorithm, due to the adaptive nature of \name.  

\subsection{Multicast Tree Construction}

We consider fully distributed algorithms for the multicast tree construction. 
Initially the tree comprises of just a single SFU which disseminates the stream to all clients. 
The choice of this initial SFU is up to the source client $c_0$ (e.g., $c_0$ can choose an SFU located close to her). 
We assume the IP address of the initial SFU is shared with all conference participants offline before the conference begins, so that the participants can connect to it and receive the stream. 
Along with the video packets, $c_0$ also sends various meta data to the SFU: sending client's identifier, number of layers sent, list of recipients clients in the session and their IP addresses, and number of layers requested by each recipient client. 

Once a tree is constructed, subsequently we allow  the SFU(s) in the tree to augment the tree by including new SFUs or removing existing SFUs from the tree. 
While augmenting a tree, each SFU can only decide its immediate (1-hop) downstream nodes in the tree. 
A downstream node for an SFU is either another SFU or a client. 
However only an SFU can forward streams; a client may only receive a stream and cannot forward it. 
While deciding the downstream nodes, the SFU must also decide the number of layers to send to each of the nodes, and the subset of clients each node (that is not a client) is responsible for distributing the stream to. 
We call these decisions as the SFU's {\em action}.
Along with the video stream, the SFU sends the following metadata to a downstream SFU: client $c_0$'s identifier, number of layers sent to the downstream SFU, list of recipients clients the downstream SFU is responsible for and their IP addresses, and number of layers requested
by each recipient client in the list. 
If a node is responsible for distributing the stream to a client $c\in C$, the node must take actions such that the node is an ancestor of $c$ in the multicast tree. 
The source $c_0$ can also change its immediate downstream SFU. 
We assume $c_0$ always sends its stream to a single SFU, which can then distribute the stream to other SFUs if needed. 
SFU and $c_0$'s actions are executed such that the video stream is received uninterrupted by the other clients at all times.  
Actions are also taken such that no cycles form in the multicast tree.
E.g., before making a connection to a new node, a node verifies that the new node is not already receiving a stream with client $c_0$ as the source. 
To alter the multicast tree, the source $c_0$ sends a special \verb|trigger_action| packet along with the video stream down the existing multicast tree. 
When a node receives a \verb|trigger_action| packet, it takes an action and forwards the \verb|trigger_action| message to its latest set of downstream nodes.  
We call each \verb|trigger_action| packet sent by $c_0$ as a round. 
Note that retaining the same set of neighbors as in the previous round is a valid action for an SFU.  

A client periodically sends a feedback to each SFU responsible for the client, on the path from $c_0$ to the client.  
The feedback contains the number of layers the client is receiving, and a specially marked packet that allows an SFU to estimate the delay in sending packets from the SFU to the client.
To estimate the delay, an SFU $s$ sends a \verb|get_delay| packet containing the SFU's identifier and the current timestamp at $s$, to all its children, who then forward the packet down the multicast tree to the clients. 
When a client receives a \verb|get_delay| packet, it immediately appends its own identifier to the packet and echos it back to $s$. 
Based on the timestamp mentioned in the received packet, and the time when the response is received by $s$, the SFU $s$ computes the delay for forwarding a packet from $s$ to the client through the multicast tree and getting a feedback response back.  
We call the computed delay as the latency between the SFU $s$ and the client. 
Based on these feedback information received from clients, SFUs decide how best to update their connections in the next round. 

\subsection{Objective}
\label{s:objective}
The objective is to construct a multicast tree which maximizes the aggregate reward (utility) of the clients. 
From the feedback received from clients, each SFU $s$ on the multicast tree computes the reward for client $c \in C_s$ as 
\begin{align} 
r_s(c) = -d(s, c) + \alpha q(c) / q^*(c), 
\end{align} where $C_s$ is the set of clients that $s$ is responsible for, $d(s, c)$ is the delay between $s$ and $c$ as estimated from the \verb|get_delay| packet and $q(c)$ is the number of layers $c$ receives.
$\alpha \geq 0$ is a user-defined parameter that dictates the trade-off between optimizing delay and video quality.
The SFU chooses its actions so that the aggregate reward $\sum_{c \in C_s} r_s(c)$ is maximized. 

\section{Motivation}

\begin{figure*}[!tbp]
  \centering
  \subfigure[]{\includegraphics[width=0.50\textwidth]{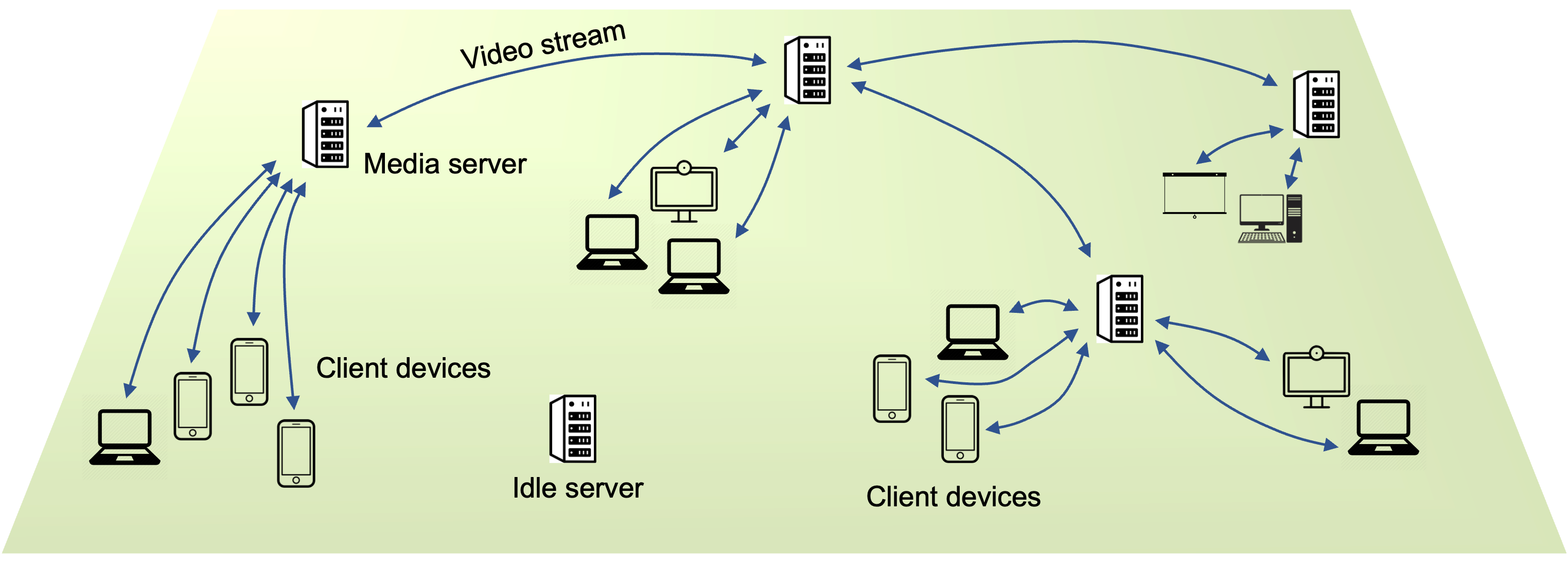}\label{fig:background}
  \vspace{15mm}}
~~
  \subfigure[]{\includegraphics[width=0.25\textwidth]{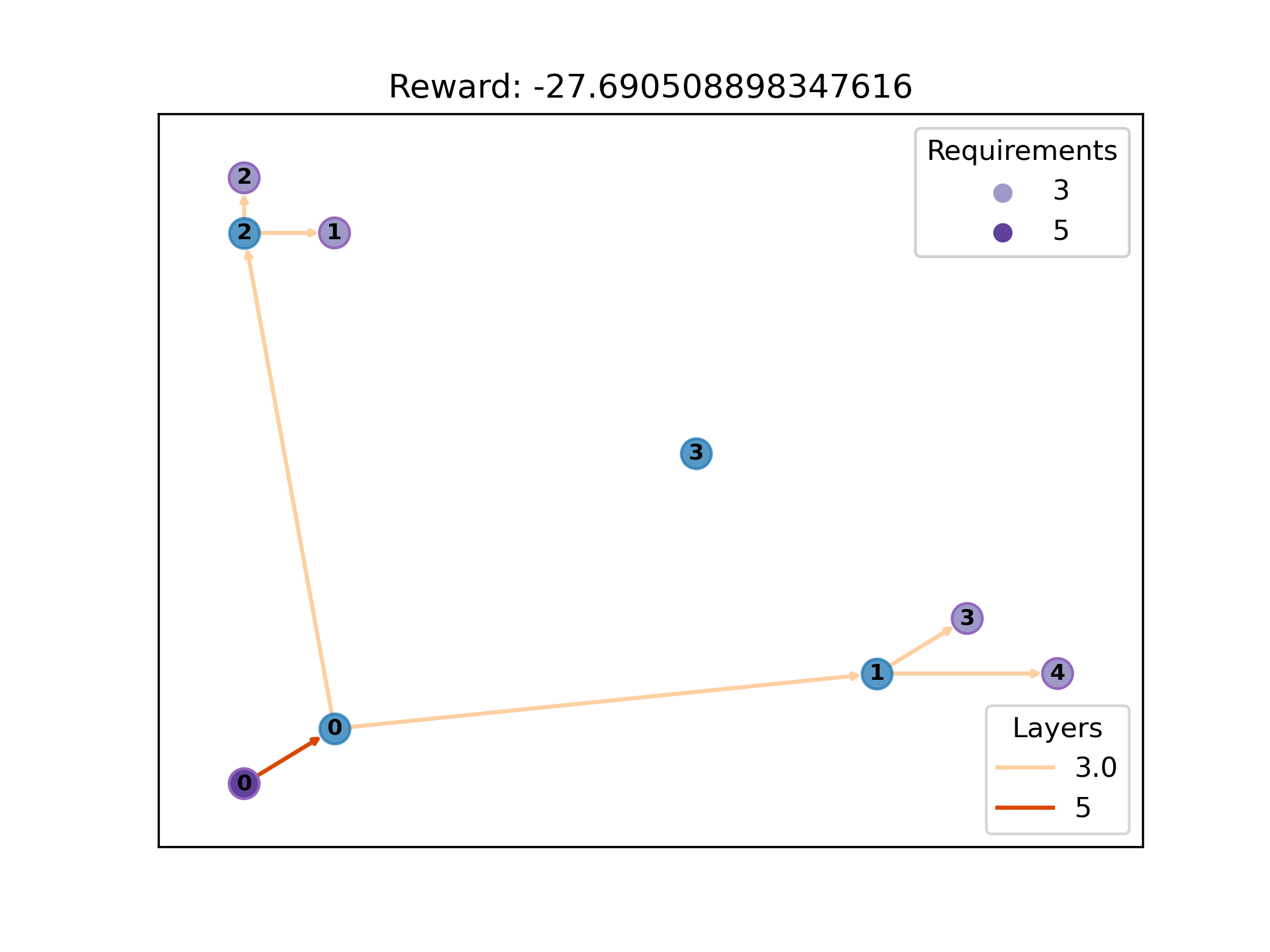}\label{fig:toy_example_optimal}}
~~
  \subfigure[]{\includegraphics[width=0.25\textwidth]{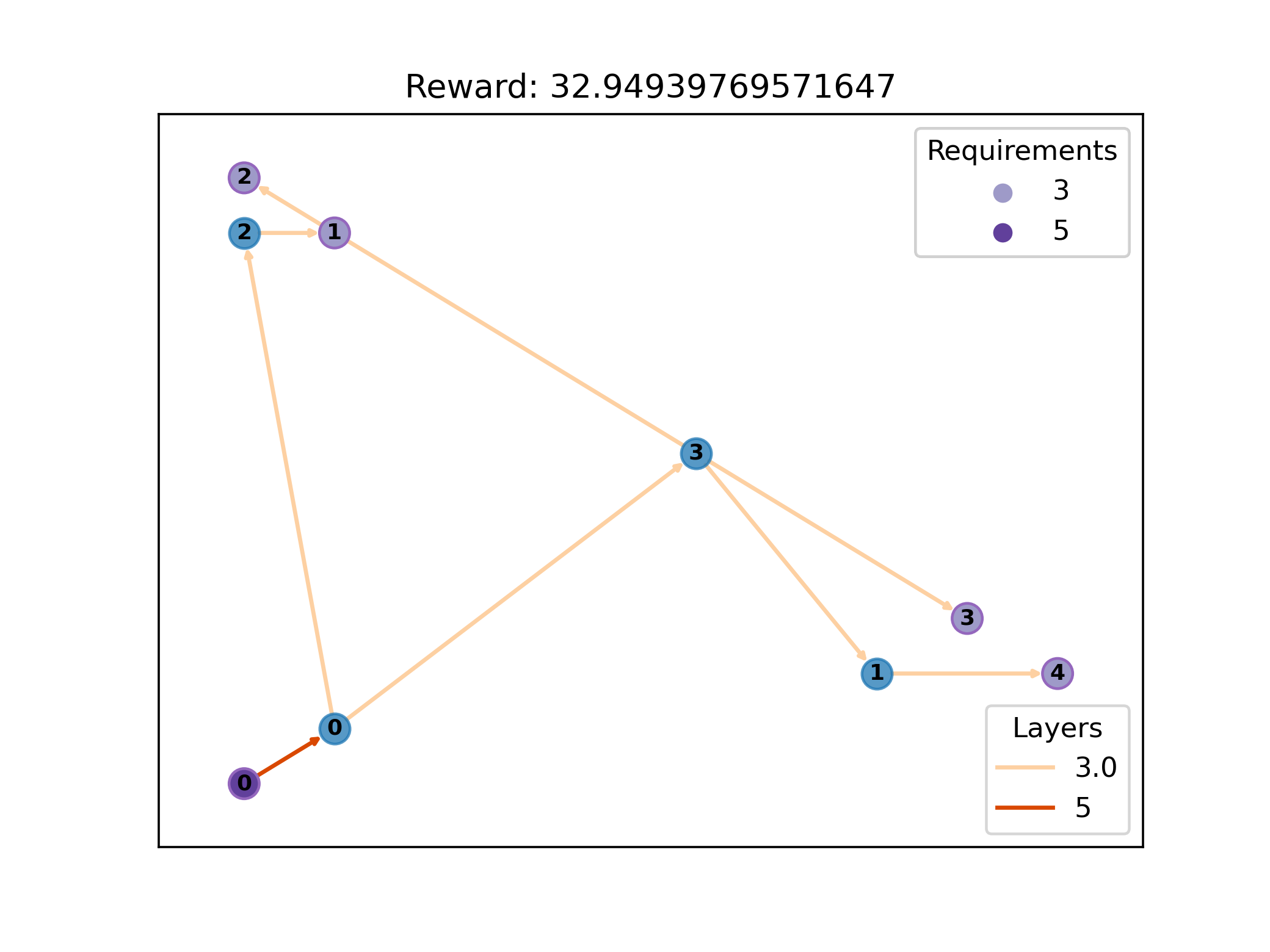}\label{fig:toy_example_non_optimal}}
  \caption{(a) \name\ considers an open, p2p network of media servers over a wide area for relaying video streams between clients. (b) An example of an optimal streaming network. (c) A sub-optimal connection compared with (b), the latency of clients 2, 3, 4 are higher.
  }
  \label{fig:backgroundandmotivatation}
\end{figure*}


To illustrate the problem and its challenges, consider the example of a conference session with 5 clients and four SFUs as shown in Figure~\ref{fig:toy_example_optimal}.
Suppose the nodes are located on a two-dimensional space with the Euclidean distance between two nodes signifying the latency of sending a packet between the nodes if they are connected to each other. 
We focus on client 0 as the source in this example. 
Each SFU has a bandwidth limit of 6 units, while each client requests to receive 3 layers of the video stream with as low a latency as possible. 
The source client generates 5 layers. 
In a centralized system, the source sends the stream to a central SFU which then forwards the stream to all other clients. 
However, with a bandwidth limit of 6 units, the central SFU cannot support high quality streams (3 layers) to all 4 clients. 
Whereas by using 3 SFUs in the multicast tree, we can deliver 3 layer streams to all 4 clients at the cost of a marginal increase in latency compared to the centralized case. 
The key challenge is how we can discover such efficient multicast tree structures through decentralized algorithms in an open setting, where comprehensive knowledge of locations of all SFUs currently online and their capacities are not known. 
Even with global knowledge, computing optimal tree structures is computationally expensive (\S\ref{s:eval}). 
In our proposed algorithm \name\ (\S\ref{s:design}), 
starting from an initial distribution tree SFUs gradually updates the tree to the state shown in Figure~\ref{fig:toy_example_optimal}. 
In this state, each client receives the full 3 layers it requested, while the latency is close to the latency of the direct link from client $0$ to the client. 
This example also illustrates that streaming via a multicast tree is bandwidth and latency efficient when there are clusters of clients located in different geographical regions.   
For clients within the same geographic region, it also illustrates the benefit of choosing SFUs from the same region. 
Figure~\ref{fig:toy_example_non_optimal} shows an example of a multicast tree that is suboptimal for this setting.  
Whereas \name\ automatically learns to ignore SFU 3 to produce the best tree configuration. 

\section{\name\ Design}
\label{s:design}
We present \name, an adaptive, decentralized algorithm to compute efficient multicast routing trees in a short amount of time. 
\name\ is motivated by the combinatorial multi-armed bandit problem (with semi-bandit feedback) applied to a decentralized setting with multiple independent agents in a shared environment~\cite{slivkins2019introduction}.  
Each agent (i.e., an SFU or $c_0$) maintains a succint model of the environment based on its past interactions with the network (past actions, and observed rewards) using which it computes the best action to take next. 
Using a model provides a convenient way to summarize observations from past interactions, without significant loss of useful information, while consuming minimal storage. 
\name\ also balances exploitative actions with exploratory actions, in which an action is randomly selected, which helps discover unseen SFU candidates with potentially good performance. 
An outline of \name\ is presented in Algorithm~\ref{algo:highlevel}.

\begin{algorithm}[t]
\DontPrintSemicolon
\SetKwInOut{Input}{input}\SetKwInOut{Output}{output}
\Input{\texttt{trigger\_action} packet from parent node, number of layers $q_\mathrm{in}(s)$ received from parent node of $s$, set of clients $C_s$ that $s$ is responsible for, IP addresses of clients in $C_s$, layer requirements of clients in $C_s$, exploration parameter $\epsilon \in (0, 1)$  
}
\Output{\texttt{action} $= \{$ set of nodes $\Gamma_s$ to forward stream to, number of layers to forward to each $s' \in \Gamma_s$, set of clients $C_{s'}$ each $s' \in \Gamma_s$ is responsible for $\}$}
\tcc{Update model based on \texttt{feedback} received since last model update.}
\texttt{model} $\leftarrow$  \textsc{UpdateModel}$(\texttt{model}, \texttt{feedback})$ \;
\tcc{Compute action balancing exploitation and exploration}
$r \leftarrow$ sample random number uniformly in $[0, 1]$ \; 
\eIf {$r \leq 1 - \epsilon$} {\texttt{action} $\leftarrow$ \textsc{EstimateBestAction}$(\texttt{model})$} 
{\texttt{action} $\leftarrow$ \textsc{RandomAction}$()$} 
Execute \texttt{action} and send \texttt{trigger\_action} to all downstream nodes $\Gamma_s \in \texttt{action}$
\caption{{\sc \namens:} Algorithm outline for computing action at SFU $s$ during a round.}
\label{algo:highlevel}
\end{algorithm}

When an SFU $s$ in the multicast tree receives a \texttt{trigger\_action} message, it uses the current model to estimate what is the best action to take. 
To encourage exploration, the best action predicted by the model is executed with probability $1-\epsilon$, where $\epsilon \in (0,1)$ is a configurable parameter. 
With probability $\epsilon$ a random action is taken. 
The model is updated each time the SFU receives a feedback from a client. 
In our experiments (\S\ref{s:eval}) we observe \name\ takes a few hundred rounds to converge to an efficient multicast tree. 
Each time the tree is reorganized, there is a potential for the video to experience disruptions at clients. 
To avoid frequent tree updates, in practice we can run \name\ in the background to discover new, efficient trees while using a stale tree to perform the actual video streaming.  
Occasionally (e.g., every 15 min), the streaming tree can be updated to the best tree found by \name\ so far. 
Another approach to minimize video disruptions is to have a setup phase lasting for a few minutes before the start of the conferencing session, during which we can run \name\ to determine an efficient multicast tree. 
Subsequently, we can fix the tree to be the best tree found during the setup phase for the entire duration of the conference. 
We leave a systematic evaluation of such implementation choices to future work. 

The main technical challenges with \name's approach in Algorithm~\ref{algo:highlevel} are (1) what model to use to summarize the environment state? (2) how to update the model with each interaction? and (3) how to compute optimal actions from the model?

\smallskip 
\noindent 
{\bf Model (\texttt{model} in Algorithm~\ref{algo:highlevel}).}
Maintaining a detailed model of the global network (e.g., where each SFU is located, their churn patterns, type of server hardware used etc.) at each agent is infeasible, as the sparse reward feedback from clients are inadequate to accurately update all model parameters. 
On the other hand, a trivial model where each candidate action is considered an independent `arm' fails to capture the rich structure in the actions which increases complexity and reduces efficiency.  

At each SFU $s$ (and client $c_0$), we consider a complete bipartite graph model 
 \( \mathcal{G}(\mathcal{S}, C_s) \) where $\mathcal{S}$ is the set of SFUs known to $s$ (including itself) and $C_s$ is the set of clients $s$ is responsible for.
 For simplicity, in our evaluations we assume $\mathcal{S} = S$, the global set of SFUs.  
 In practice, if the size of $\mathcal{S}$ is large, heuristics may be used to prune down the set of candidate SFUs considered. 
 Each edge $(s', c') \in \mathcal{G}$ has two weights associated with it: $\hat{d}(s, s',c')$ which is an estimate of the latency between $s$ and $c'$ if the stream is forwarded through $s'$ (and $s'$ is responsible for delivering the stream to $c'$), and $\hat{\phi}(s',c')$ which is an estimate of the average number of layers $c'$ receives if $s$ forwards the stream to $s'$ (and $s'$ is responsible for delivering the stream to $c'$) per unit layer sent to $s'$.
 In other words, if $s$ sends $q(s, s')$ layers to $s'$ and makes $s'$ responsible for $c'$, we estimate $q(s, s') \times \hat{\phi}(s',c')$ as the number of layers $c'$ receives. 
 Maintaining an estimate $\hat{\phi}(s', c')$ of the number of layers received by $c'$ per unit layer sent to $s'$ allows us to estimate the number of layers received by $c'$ based on the action $q(s, s')$ at $s'$. 

\smallskip
\noindent 
{\bf Computing the best action (\textsc{EstimateBestAction()} in Algorithm~\ref{algo:highlevel}).}
An action at SFU $s$ can be represented using two functions: a client assignment map $a(s, \cdot):\mathcal{C} \rightarrow \mathcal{S}$ and number of layers forwarded $q(s, \cdot):\mathcal{S}\cup \mathcal{C} \rightarrow \mathbb{N}$ where $\mathbb{N} = \{0,1,2,\ldots\}$. 
If $a(s, c') = s'$, it means $s$ forwards a non-zero number of layers to $s'$ while also informing $s'$ to be responsible for $c'$. 
If $q(s, s') > 0$ for $s' \in \mathcal{S}$, it means SFU $s$ forwards $q(s, s')$ layers to $s'$. 
Similarly, if $q(s, c') > 0$ for $c'\in \mathcal{C}$, it means $s$ forwards $q(s, c')$ layers directly to $c'$. 
For consistency, we must have $q(s, s')>0$ iff $\exists~ c' \in \mathcal{C}$ such that $a(s, c') = s'$ for any $s' \in \mathcal{S}$.  
Similarly, $q(s, c') > 0$ iff $a(s, c') = s$ for any $c' \in \mathcal{C}$. 

For any candidate action $(a(s, \cdot), q(s, \cdot))$ where $a, q$ are consistent with each other as noted above, we can estimate the reward incurred by the action as 
\begin{align}
\sum_{c' \in C_s} \left( -\hat{d}(s, a(s, c'), c') + \alpha \mathbf{1}_{a(s,c') \neq s} \frac{\min(q(s, a(s, c')) \hat{\phi}(a(s, c'), c'), q(c'))}{q(c')} \right. \notag \\
\left. + ~\alpha \mathbf{1}_{a(s,c') = s} \frac{\min(q(s, c'), q(c'))}{q(c')} \right) \label{eq:estreward}
\end{align}
following our reward model (\S\ref{s:objective}). 
The \textsc{EstimateBestAction()} outputs the action with the highest estimated reward. 
A na\"ive method to find the best action is to exhaustively compute the reward for all possible actions and choose the best one. 
However, the set of possible actions grows prohibitively large in size even for moderate sized problem instances. 
We use an integer-quadratic-program (IQP) to compute the best action as follows: 
\begin{align}
\max \quad \sum_{c' \in C_s} & \left( - \sum_{s' \in \mathcal{S} } x(s', c') \hat{d}(s, s', c') + \sum_{\substack{s' \in \mathcal{S} \\ s' \neq s}} \alpha x(s', c') \frac{q(s, s', c')}{q(c')} + \alpha x(s, c') \frac{q(s, c')}{q(c')} \right) \\
\text{such that } \quad &x(s', c') \in \{0, 1\} \quad \forall s' \in \mathcal{S}, c' \in C_s \\
&q(s, s') \in \mathbb{N} \quad \forall s' \in \mathcal{S}, s' \neq s \\
&q(s, s', c') \in \mathbb{N} \quad \forall s' \in \mathcal{S}, s' \neq s, c' \in C_s \\ 
&q(s, c') \in \mathbb{N} \quad \forall c' \in C_s \\
&q(s, s', c') = \min(q(s, s')\hat{\phi}(s', c'), q(c')) \quad \forall s' \in \mathcal{S}, s' \neq s, c' \in C_s \label{const:numlayersent} \\
&q(s, c') \leq q(c') \quad \forall c' \in C_s \label{const:numlayersentclient} \\
& \sum_{\substack{s' \in \mathcal{S} \\ s' \neq s}} q(s, s') + \sum_{c' \in C_s} q(s, c') \leq \min(b_s, q_\mathrm{in}(s)). \label{const:capacity}
\end{align}
Here, $x(s', c')$ is an indicator variable that denotes whether client $c'$ is mapped to SFU $s$; 
$q(s, s')$ denotes the number of layers $s$ sends to a downstream neighbors $s'$; 
$q(s, s', c')$ is an estimate of the number of layers client $c'$ receives if the stream for $c'$ is sent through $s'$; 
$q(s, c')$ is the number of layers $s$ sends to client $c'$ if $s$ forms a direct connection to $c'$.
Eq.~\eqref{const:numlayersent} says that if $q(s, s')$ layers are sent by $s$ to $s'$ and $s'$ is responsible for $c'$, then the number of layers received by $c'$ is at most $q(s, s')\hat{\phi}(s', c')$ and at most $q(c')$ (number of layers requested by $c'$). 
Similarly Eq.~\eqref{const:numlayersentclient} says the number of layers sent by $s$ directly to a client $c'$ cannot exceed client $c'$'s demand $q(c')$. 
Finally Eq.~\eqref{const:capacity} says the total number of layers sent by $s$ cannot exceed the bandwidth limit of $s$. 
Note that despite the non-linear min function in Eq.~\eqref{const:numlayersent}, the constraint can be converted in to a linear constraint using auxiliary variables. 
In our experiments, we observe the above IQP finds the exact optimal solution for small to moderate sized problem instances with few tens of nodes (see \S\ref{s:eval}). 
For larger networks, we run the above optimization with a time cutoff specified.


\smallskip 
\noindent 
{\bf Updating the model (\textsc{UpdateModel()} in Algorithm~\ref{algo:highlevel}).}
The edge weight parameters in our bipartite graph model must be estimated, as SFU $s$ may not have exact information on downstream nodes' locations or their subtree topologies.
When a client or SFU is first included in the bipartite graph model $\mathcal{G}$, the latency and quality parameters are initialized to zero. 
Each time an action is played, the SFU updates its model using the feedback received from the clients as follows. 
If the stream for client $c' \in C_s$ is routed through SFU $s' \in \mathcal{S}$, we have 
\begin{align}
\hat{d}(s, s', c') &= \hat{d}(s, s', c') + \eta (d(s,c') - \hat{d}(s, s', c')) \\
\hat{\phi}(s', c') &= \hat{\phi}(s', c') + \eta' (\frac{q(c')}{q(s, s')} - \hat{\phi}(s', c')),
\end{align}
where $\eta, \eta'$ are step size parameters. 
Similarly, for a client $c' \in C_s$ that is directly sent the stream from $s$, we have 
\begin{align}
\hat{d}(s, c') &= \hat{d}(s, c') + \eta (d(s,c') - \hat{d}(s, c')) \\
\hat{\phi}(s, c') &= 1.  
\end{align}
To encourage exploration, we also include an additive bonus term to the edge weights based on the number of times the edge has been selected~\cite{auer2002finite}.   
It is possible for a feedback message from a client to take a long time to be received by SFU $s$, due to inefficient, long paths to the client from $s$.  To prevent stalling, we therefore set a deadline and update the model parameters using default target values if a feedback is not received by the deadline. 


\smallskip 
\noindent 
{\bf Increasing stability.}
Depending on the depth of an SFU in the multicast tree, it can take several rounds to evaluate the true quality of the SFU's actions. 
E.g., consider an SFU $s$ that has 3 levels of other SFUs below it in the subtree rooted at the $s$. 
SFUs in each of those levels must perform sufficient exploration and exploitation rounds in their respective action spaces to determine an efficient topology for the subtree. 
Only when this subtree converges to an efficient topology, does the true impact of SFU $s$'s action becomes known. 
The higher up SFU $s$ is in the subtree, the greater could be the number of rounds necessary to assess the true quality of $s$'s action. 


To avoid underestimating the performance of an action, we consider a variant of our proposed algorithm when an SFU makes a new action only after a fixed number of rounds have elapsed since the last action.
This interval between successive actions is decided based on the depth of the subtree rooted at the SFU with deeper subtrees requiring a longer interval. 



\section{Evaluation}
\label{s:eval}

\subsection{Simulator Design}

\begin{figure}[t]
    \centering
    \includegraphics[width=0.7\textwidth]{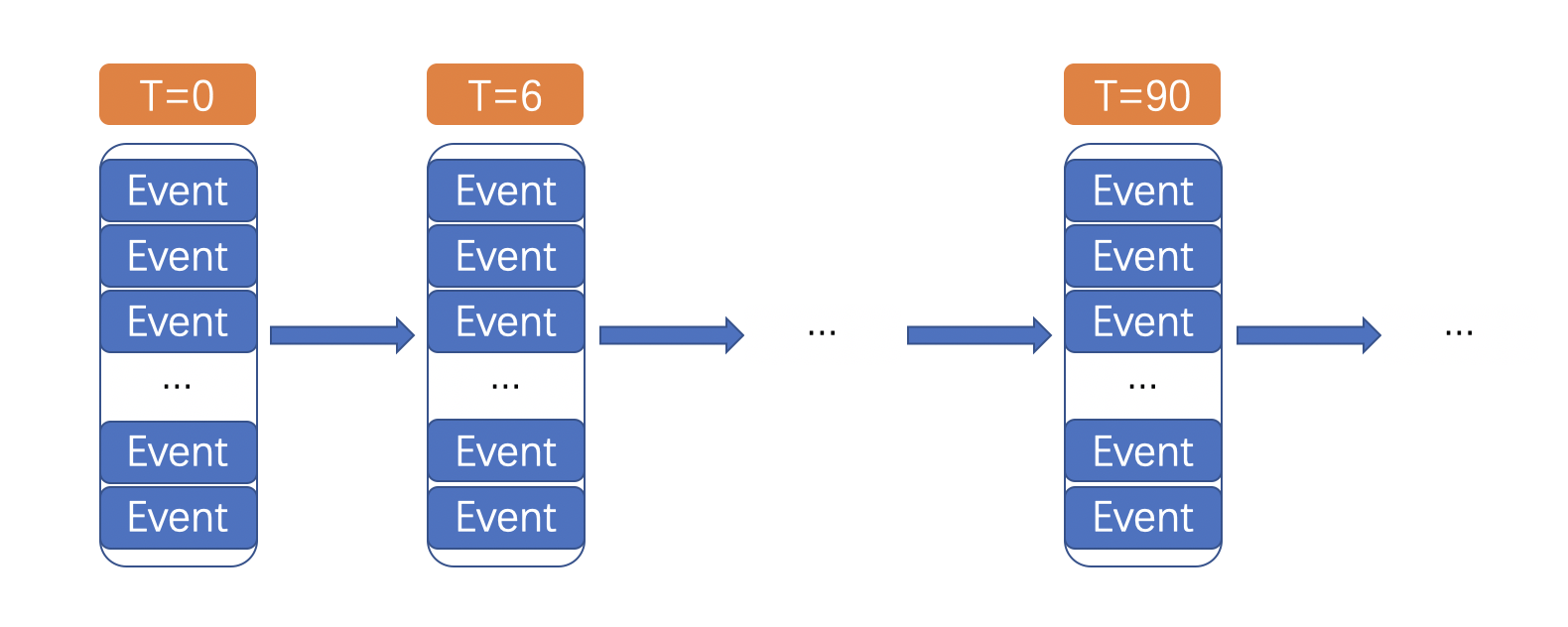}
    \caption{Diagram of the Event Queue Model. }
    \label{fig:simulator_intro}
\end{figure}


We evaluate \name\ on a custom event-based simulator written in Python.\footnote{The simulator will be made publicly available in the final version of the paper.} 
Each packet (message) sent or received by a node in the network is an event.
Events are scheduled to occur following the network's stipulated link delays using a linked list of queues datastructure (Figure~\ref{fig:simulator_intro}). 
 Each node of the linked list is associated with a time stamp and an event queue. 
 The nodes are ordered by the time stamp. An event is first scheduled by adding the event to the event queue and then evaluated when the event queue is being evaluated.
The nodes of events are evaluated based on the order of the nodes, which is the chronological order. 
A new event could be triggered when evaluating the current event. For example, when evaluating the event ‘client receives a packet’, a new event, ‘client sends a feedback’ is triggered. The new event is inserted into the linked list based on the time it will be evaluated.


\subsection{Experiment Setup}

We consider a p2p network of clients and SFUs, with each node assumed to be located at a point on a two-dimensional plane.  For any two nodes, the latency of sending a packet between the nodes is set to be proportional to the Euclidean distance between the nodes on the 2D plane. 
We vary the number of clients between 5--50 and the number of SFUs between 3--10 in the experiments. 
The bandwidth of each SFU is chosen such that multiple SFUs are required in a multicast tree to achieve good QoS. 
In practice, a single SFU can support over 30 participants in a session with a 500 Mbps link~\cite{Jitsiperformeval}. 
However, if there are many concurrent conference sessions served by the same SFU, the amount of available bandwidth to any one conference diminishes.  
Each client has a quality (number of layers) requirement as discussed in \S\ref{s:model}. 
Every 5s, a client initiates a \texttt{trigger\_action} command which is propagated down its multicast tree trigerring actions at the SFUs.  
We use a step size proportional to the number of SFUs involved in each setting, and we set $\alpha=0.7$. 
Each time when the algorithm is activated, it is recorded as one round. The model converges in a short amount of time in all experiment settings and remains stable after the converge point.

\smallskip 
\noindent 
{\bf Baselines.}
We compare \name's QoE against the global optimum computed using integer programming.
The global optimum is a centralized scheme that takes complete information about the SFUs and clients locations, SFU bandwidth limits and client QoE requirements as input to compute the best multicast trees and routing. 
Constraining the distribution paths to be a tree introduces an exponential number of constraints in the integer program. 
Thus, we are able to compute the global optimum only for small networks. 
The complete integer programming formulation has been presented in Appendix~\ref{s:ipbaseline}. 
We use Gurobi Optimization Toolbox to solve this problem. A time cut-off of 300s is set for large settings (10 SFUs, 50 clients).

We also consider a nearest-server baseline, in which the source client selects a server that is closest to it, to which all other clients connect forming a star topology. 
Many cloud-based conferencing application follow this policy. 

\smallskip 
\noindent 
{\bf Conference configurations.}
We consider various node location settings to illustrate the QoE provided by \name\ and its convergence behavior:  

\smallskip
\noindent 
{\em (1) Structured node locations.} 
In this setting, we place clients and SFUs at carefully chosen locations such that it is intuitively clear what the optimal multicast tree should be. 
Specifically, we arrange the nodes as a tree (placed over the 2D plane) with client 0 as the root, the SFUs as interior nodes and all other clients as leaves. 
We evaluate whether \name\ automatically discovers this tree. 
The global optimum baseline also computes the planted trees as optimum. 
We consider 3 settings representing small (3 SFUs, 5 clients), medium (7 SFUs, 11 clients) and large (10 SFUs, 50 clients) conference scenarios. 

\smallskip
\noindent 
{\em (2) Random node locations.}
Next, we consider a setting where nodes are randomly placed on the 2D plane. 
As before we evaluate the small, medium and large conference sizes. 
The random node location setting is designed to mimic the distribution of nodes in real-world wide-area p2p networks. 

\smallskip 
\noindent 
{\em (3) Real-world locations.}
To compare against popular centralized conferencing system today (e.g., Zoom), we consider a setting where clients and SFUs are located at various major cities around the world, with round-trip-times between cities obtained from a ping measurement dataset~\cite{globalpingdata}.  
We also include Zoom data center locations~\cite{zoomdclocations} to simulate Zoom's stream forwarding policy. 

\subsection{Results}

\name’s QoS is comparable to the global optimum baseline in all the three scenarios. 
The time \name\ takes to converge depends on the size of setting and model hyper-parameters; a small or medium setting converges in 100 to 200 rounds, while for large settings it takes 300 to 500 rounds to converge.

\begin{figure*}[!tbp]
  \centering
  \subfigure[]{\includegraphics[width=0.5\textwidth]{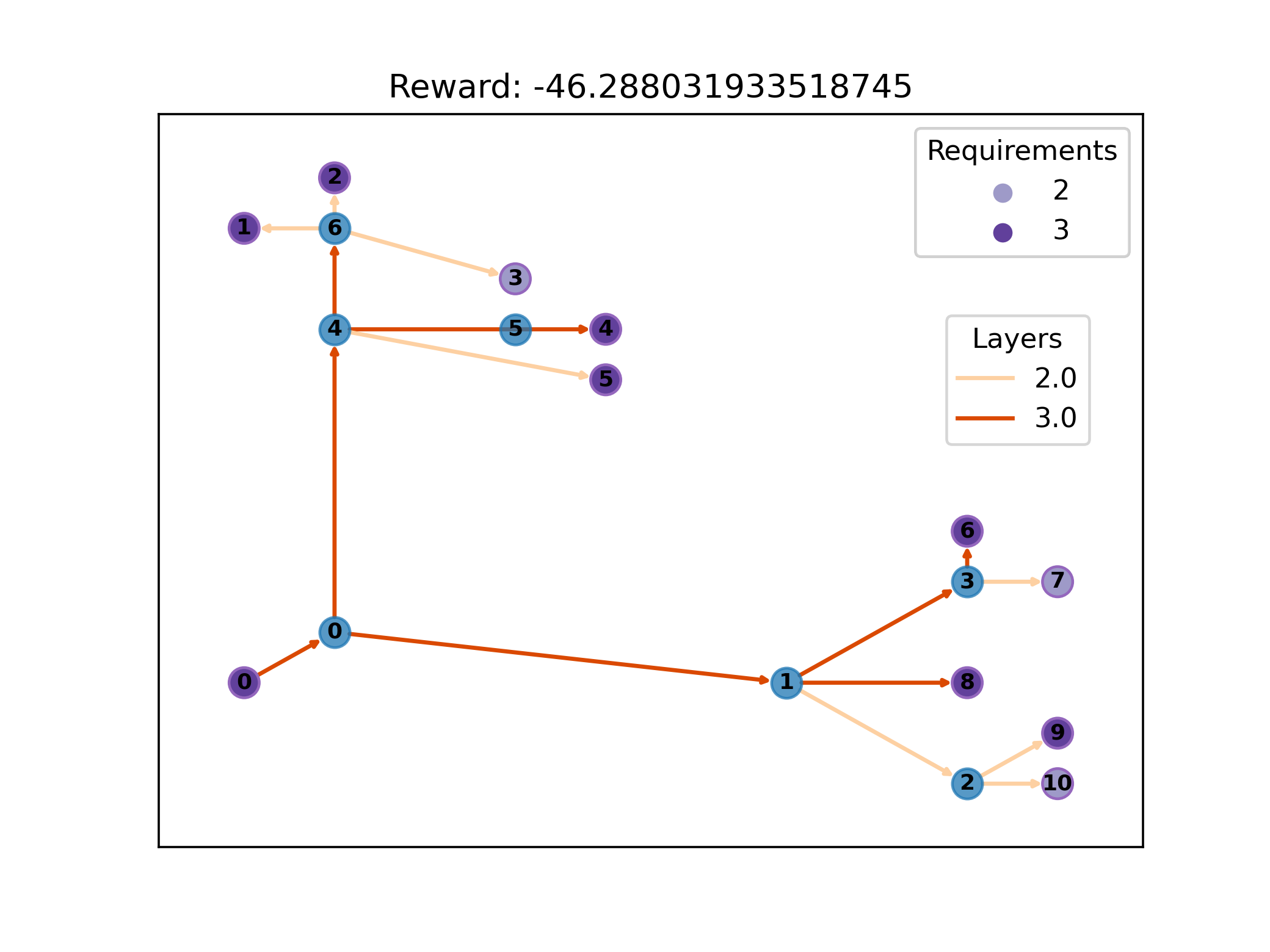}\label{fig:strc_m_1}}
~~
  \subfigure[]{\includegraphics[width=0.5\textwidth]{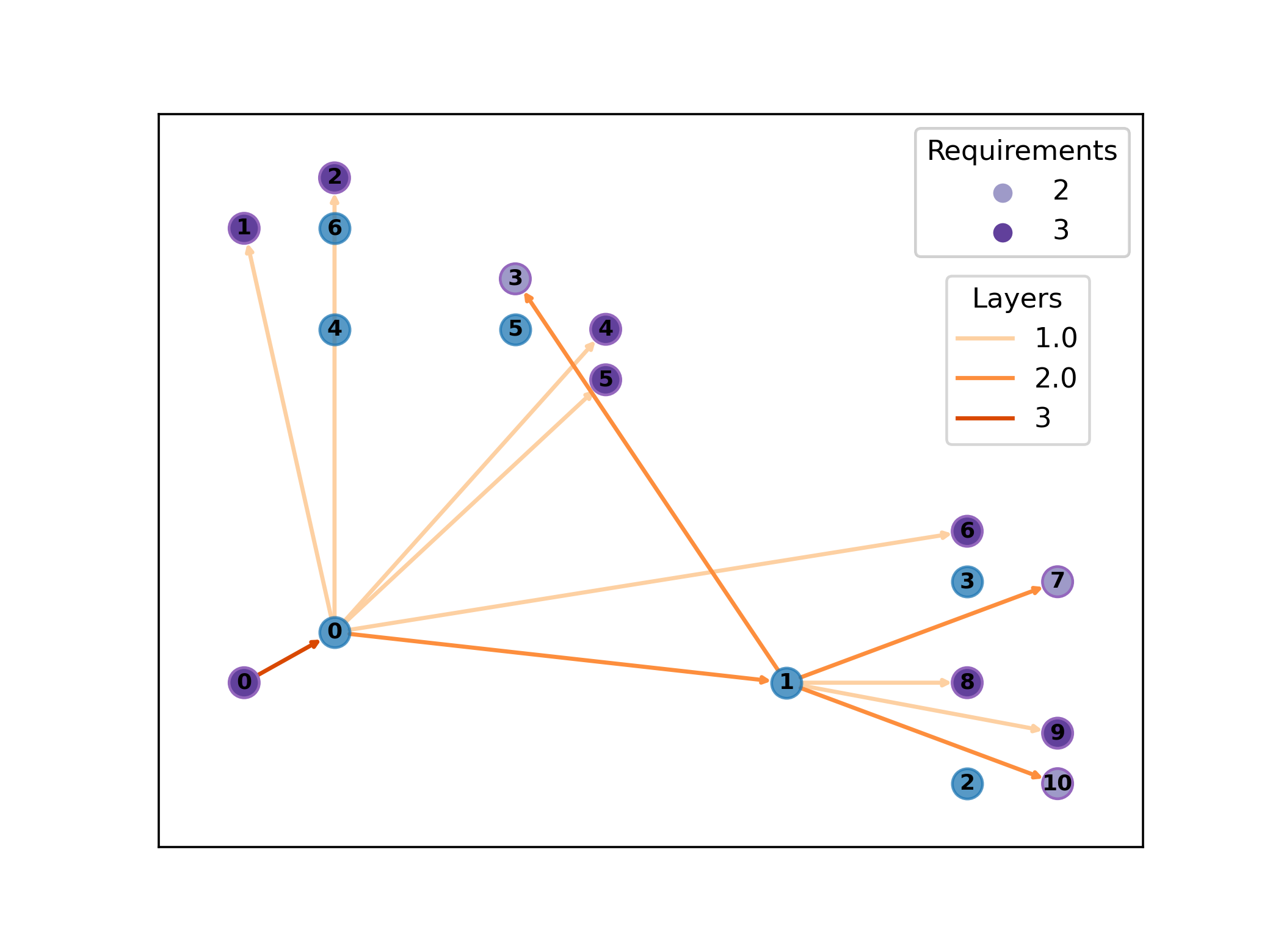}\label{fig:strc_m_2}}
~~
  \caption{Connection of 7 SFUs, 11 clients structured setting. (a) Connection generated by \name. (b) Connection generated by the Baseline.
  }
  \label{fig:strc_m}
\end{figure*}

\begin{figure*}[!tbp]
  \centering
  \subfigure[]{\includegraphics[width=0.5\textwidth]{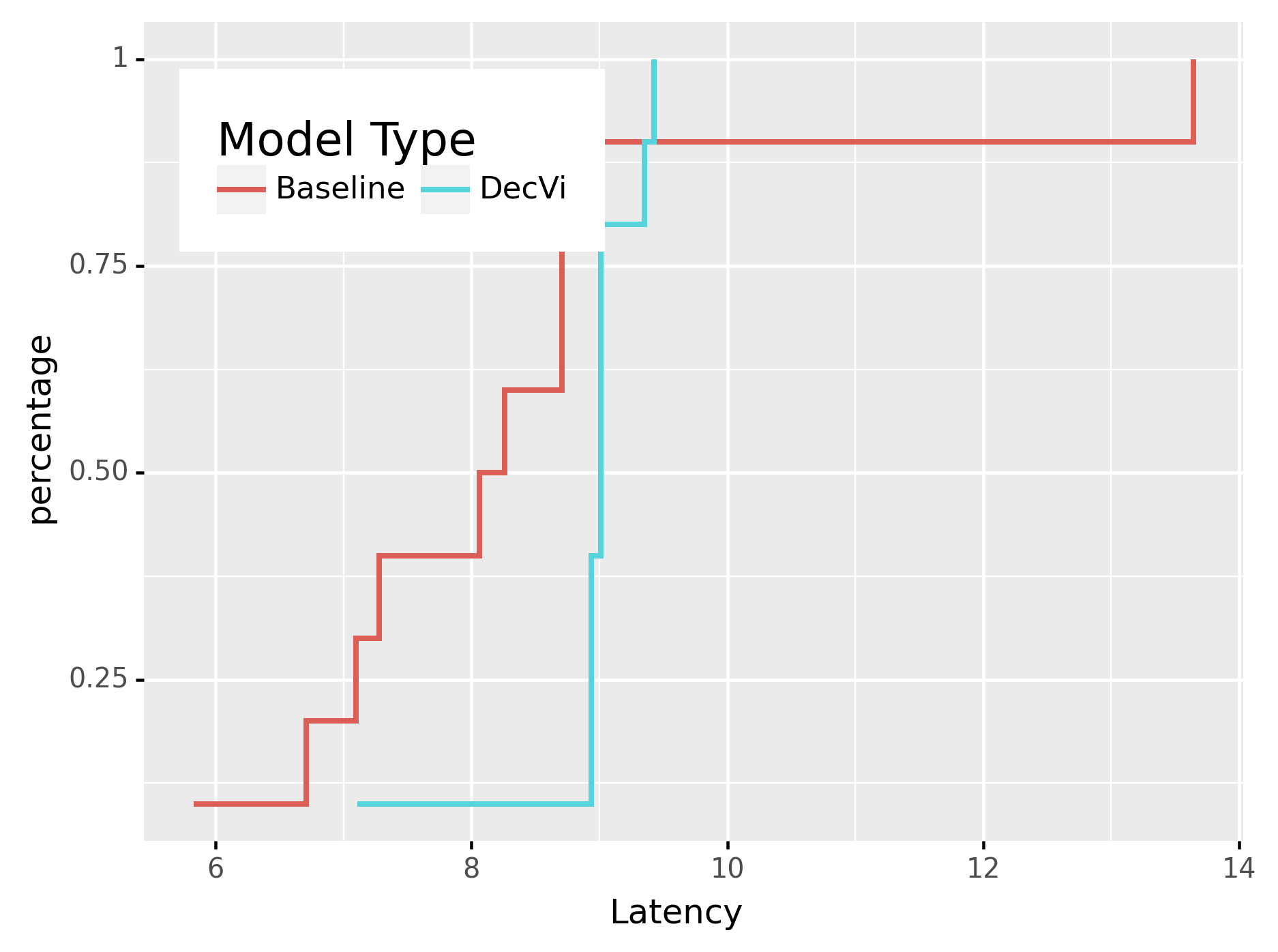}\label{fig:strc_m_cdf_1}}
~~
  \subfigure[]{\includegraphics[width=0.5\textwidth]{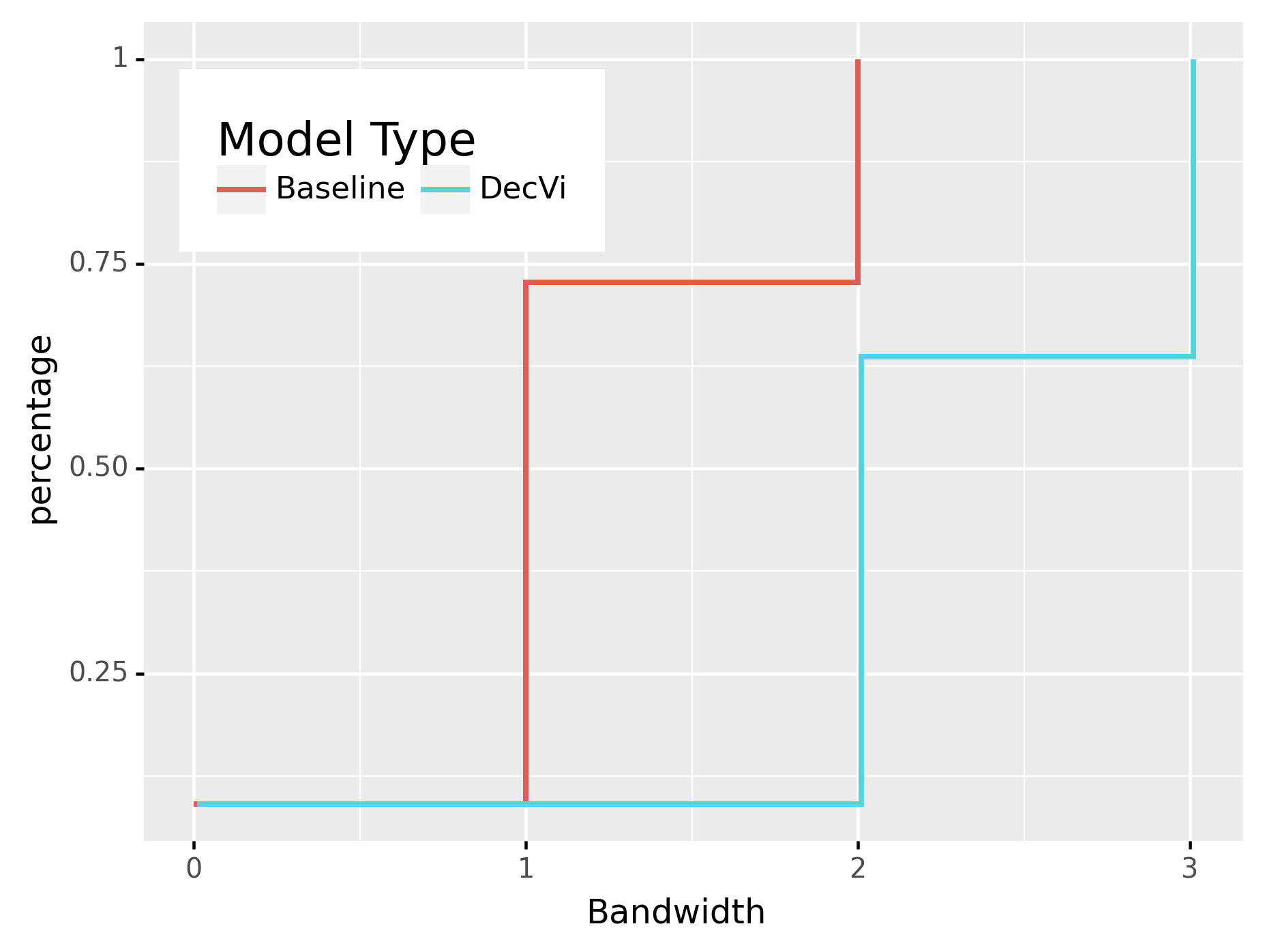}\label{fig:strc_m_cdf_2}}
~~
  \caption{CDF comparison between \name\ and baseline under the 7 SFUs, 11 clients structured setting. (a) CDF of latency. (b) CDF of bandwidth.
  }
  \label{fig:strc_m_cdf}
\end{figure*}

\begin{figure*}[!tbp]
  \centering
  \subfigure[]{\includegraphics[width=0.5\textwidth]{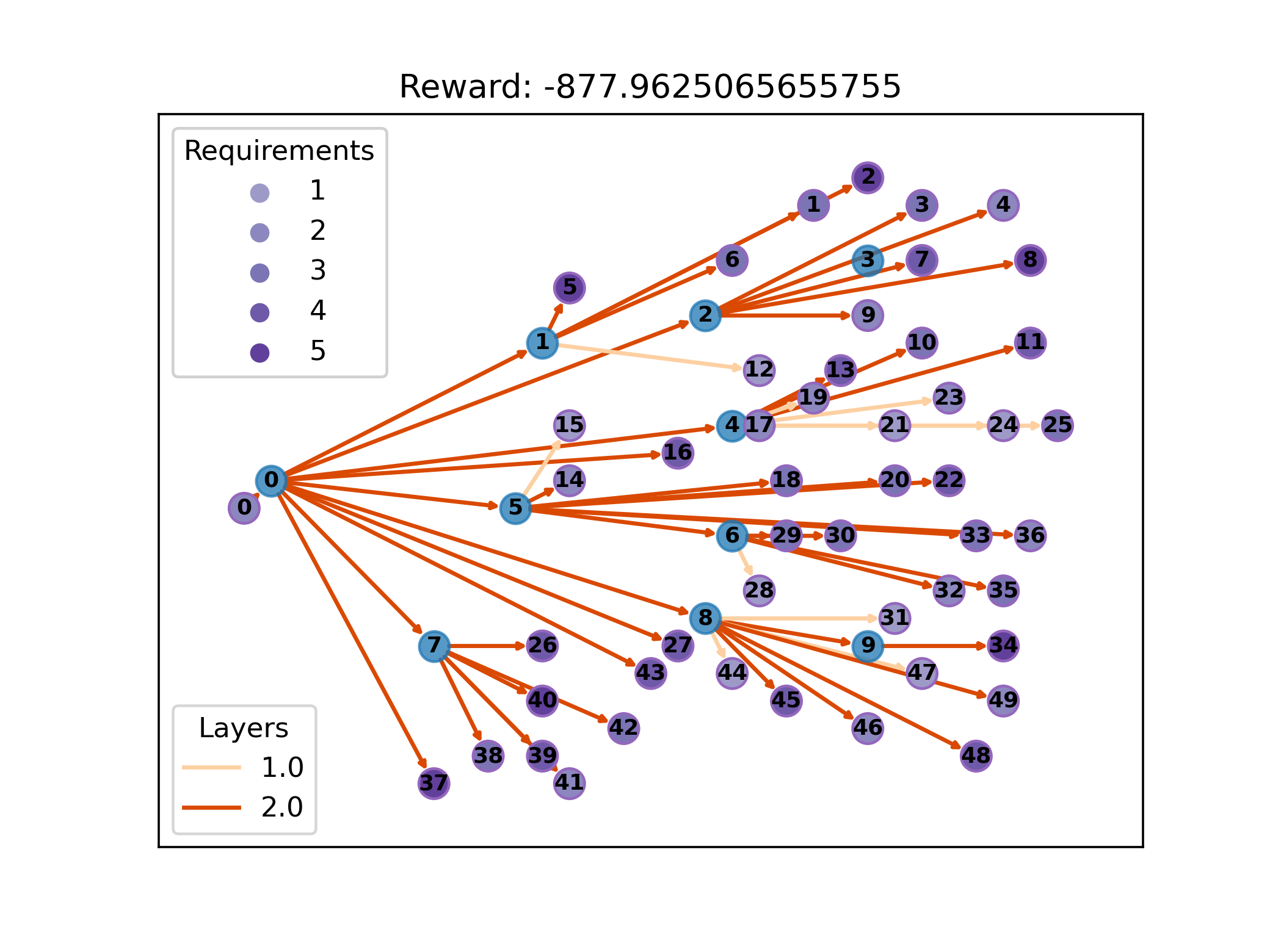}\label{fig:strc_l_1}}
~~
  \subfigure[]{\includegraphics[width=0.5\textwidth]{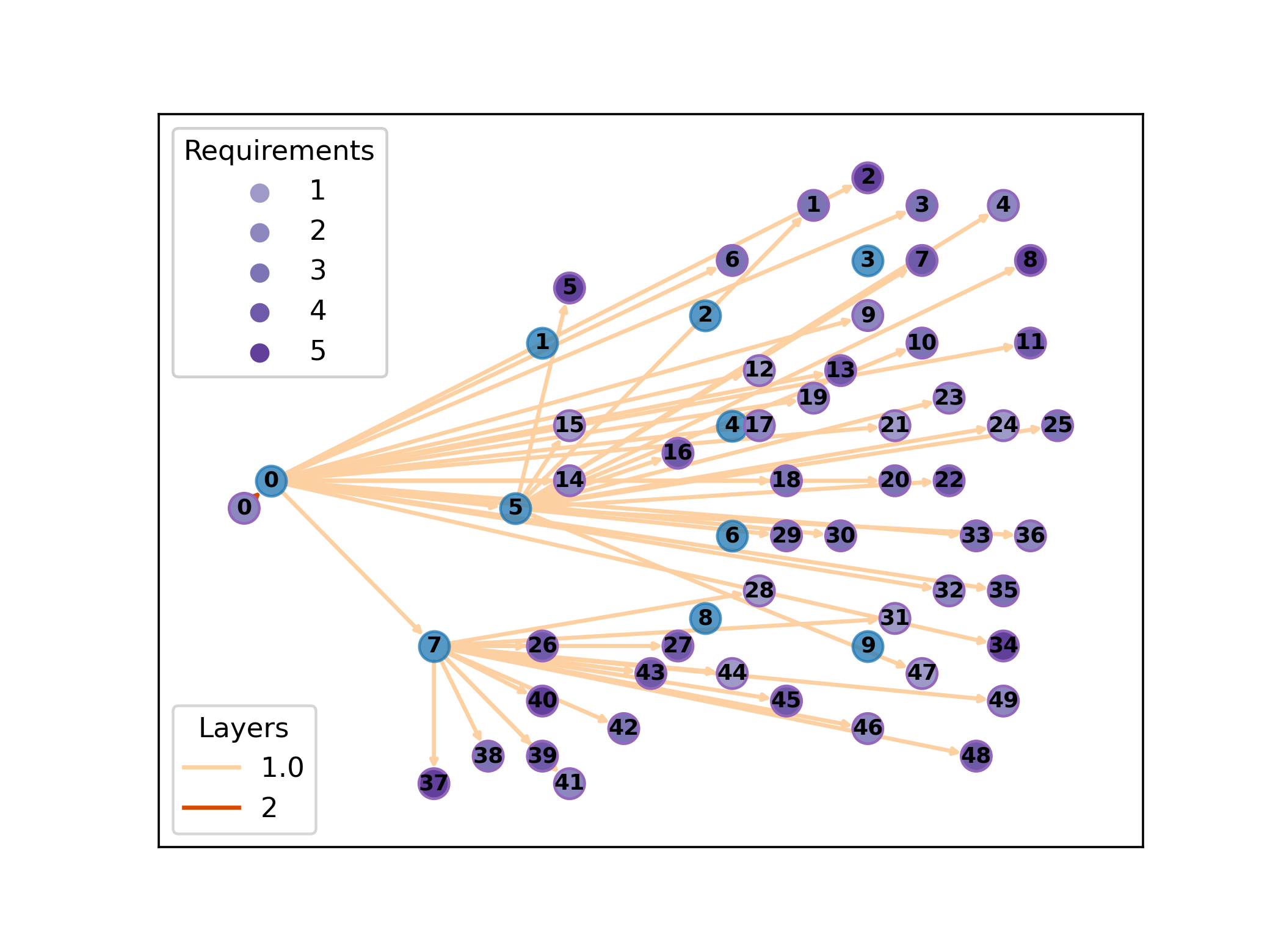}\label{fig:strc_l_2}}
~~
  \caption{Connection of 10 SFUs, 50 clients structured setting. (a) Connection generated by \name. (b) Connection generated by the baseline.
  }
  \label{fig:strc_l}
\end{figure*}

\begin{figure*}[!tbp]
  \centering
  \subfigure[]{\includegraphics[width=0.5\textwidth]{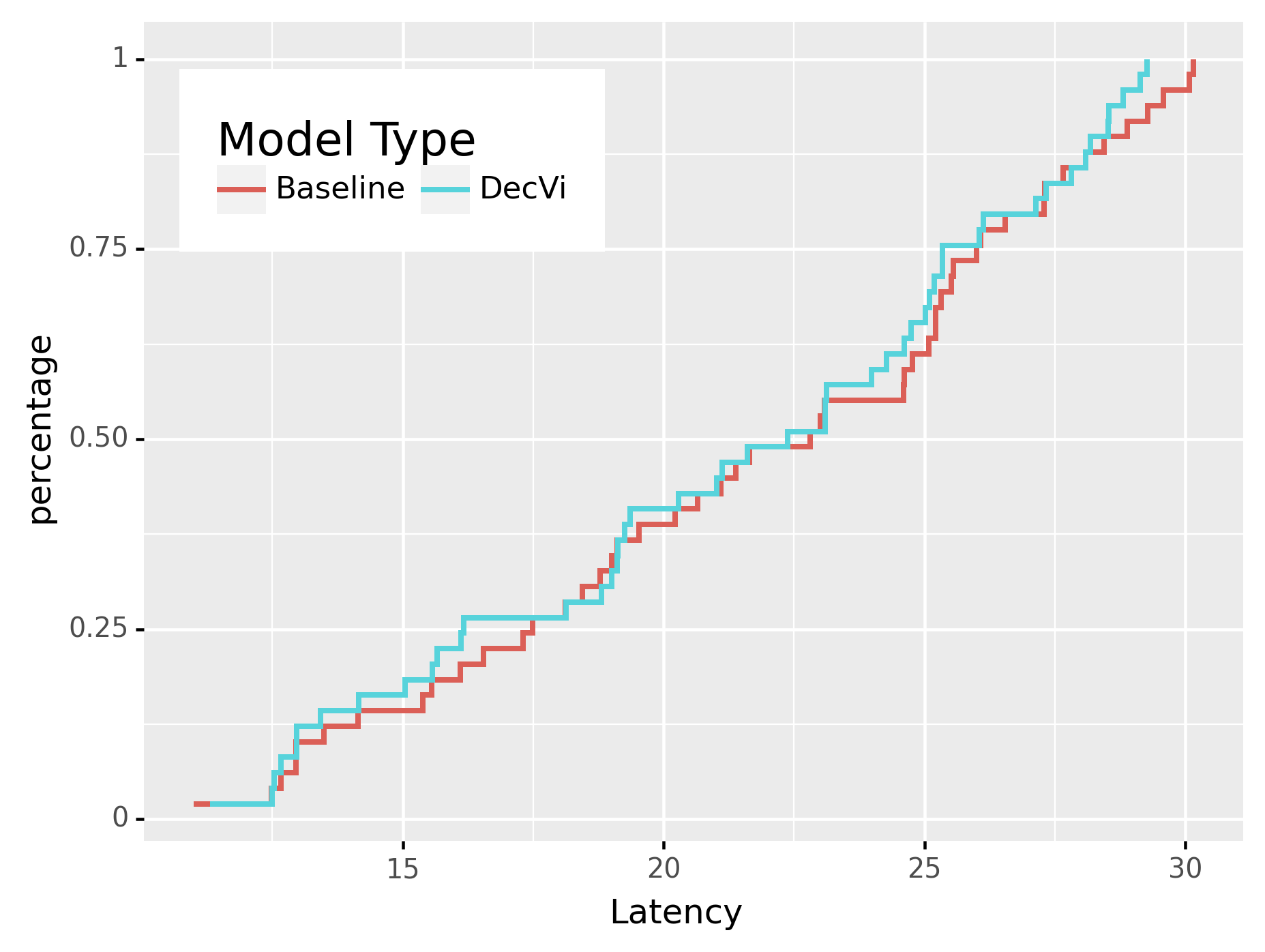}\label{fig:strc_l_cdf_1}}
~~
  \subfigure[]{\includegraphics[width=0.5\textwidth]{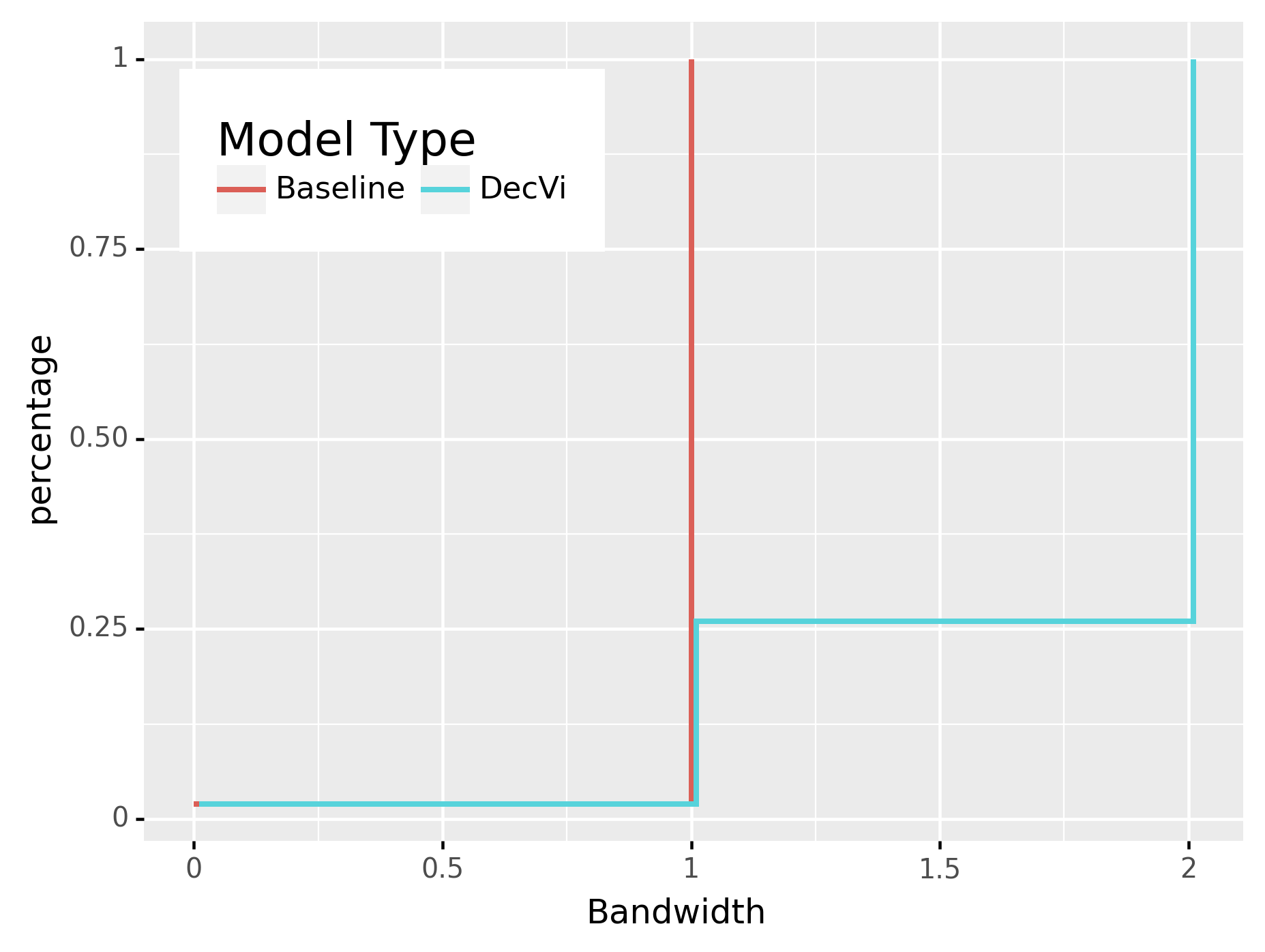}\label{fig:strc_l_cdf_2}}
~~
  \caption{CDF comparison between \name\ and baseline under the 10 SFUs, 50 clients structured setting. (a) CDF of latency. (b) CDF of bandwidth.
  }
  \label{fig:strc_l_cdf}
\end{figure*}

\subsubsection{Structured Settings}
Figure \ref{fig:strc_m} and \ref{fig:strc_l} show the connections of structured settings. Figure \ref{fig:strc_m_1} and \ref{fig:strc_l_1} show the connection given by the proposed model, Figure \ref{fig:strc_m_2} and \ref{fig:strc_l_2} show the connection given by the global optimum baseline, respectively. The connections generated by \name\ closely match the optimal connections. The shade of edge color represents the number of layers actually sent through that streaming connection. Figure \ref{fig:strc_m_cdf} and \ref{fig:strc_l_cdf}  show the comparison of the latency and bandwidth of each client between \name\ and the baseline, given by CDF.

Performance comparison between \name\ and the baseline are shown by CDF of latency and bandwidth respectively. The CDF of latency comparison, which is in (a), shows that \name\ has a comparable latency QoS to the baseline.

\subsubsection{Connections under Different QoS Preferences}
\begin{figure*}[!tbp]
  \centering
  \subfigure[]{\includegraphics[width=0.5\textwidth]{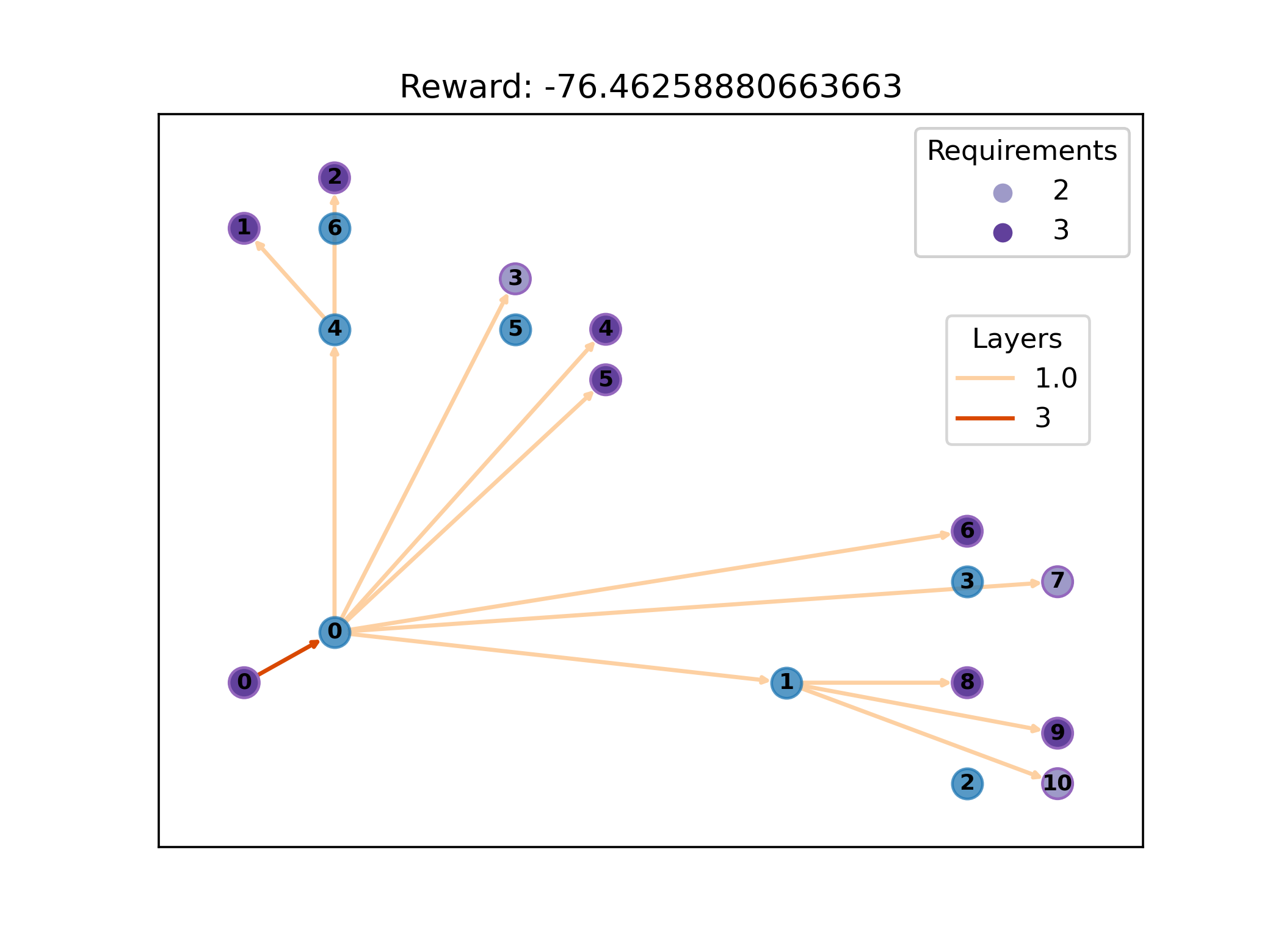}\label{fig:strc_m_qos0}}
~~
  \subfigure[]{\includegraphics[width=0.5\textwidth]{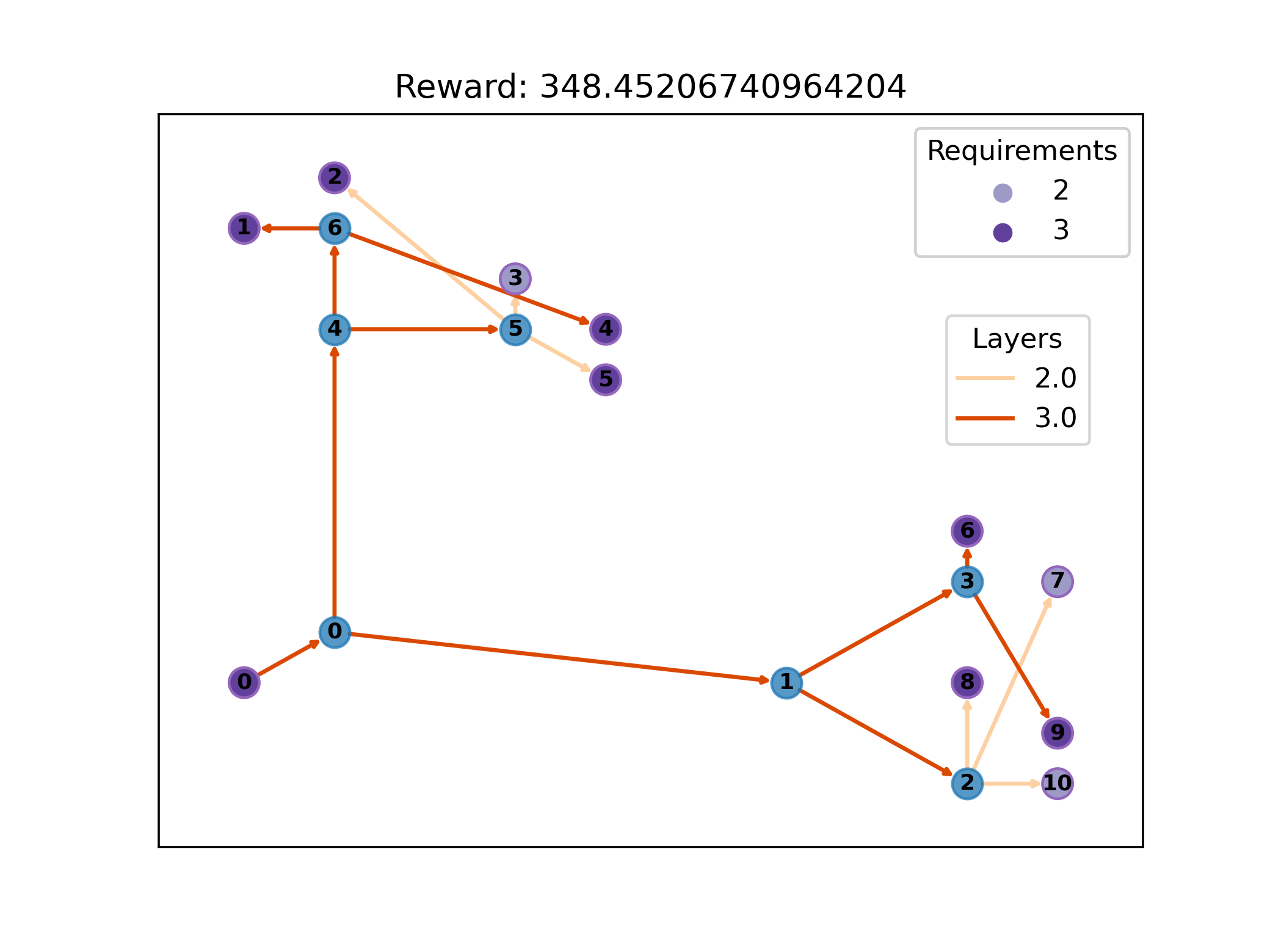}\label{fig:strc_m_qos50}}
~~
  \caption{\name's performance on 7 SFUs, 11 clients structured setting under different QoS parameters. (a) $\alpha=0$. (b) $\alpha=50$.}
  \label{fig:strc_m_qos}
\end{figure*}

\begin{figure*}[!tbp]
  \centering
  \subfigure[]{\includegraphics[width=0.5\textwidth]{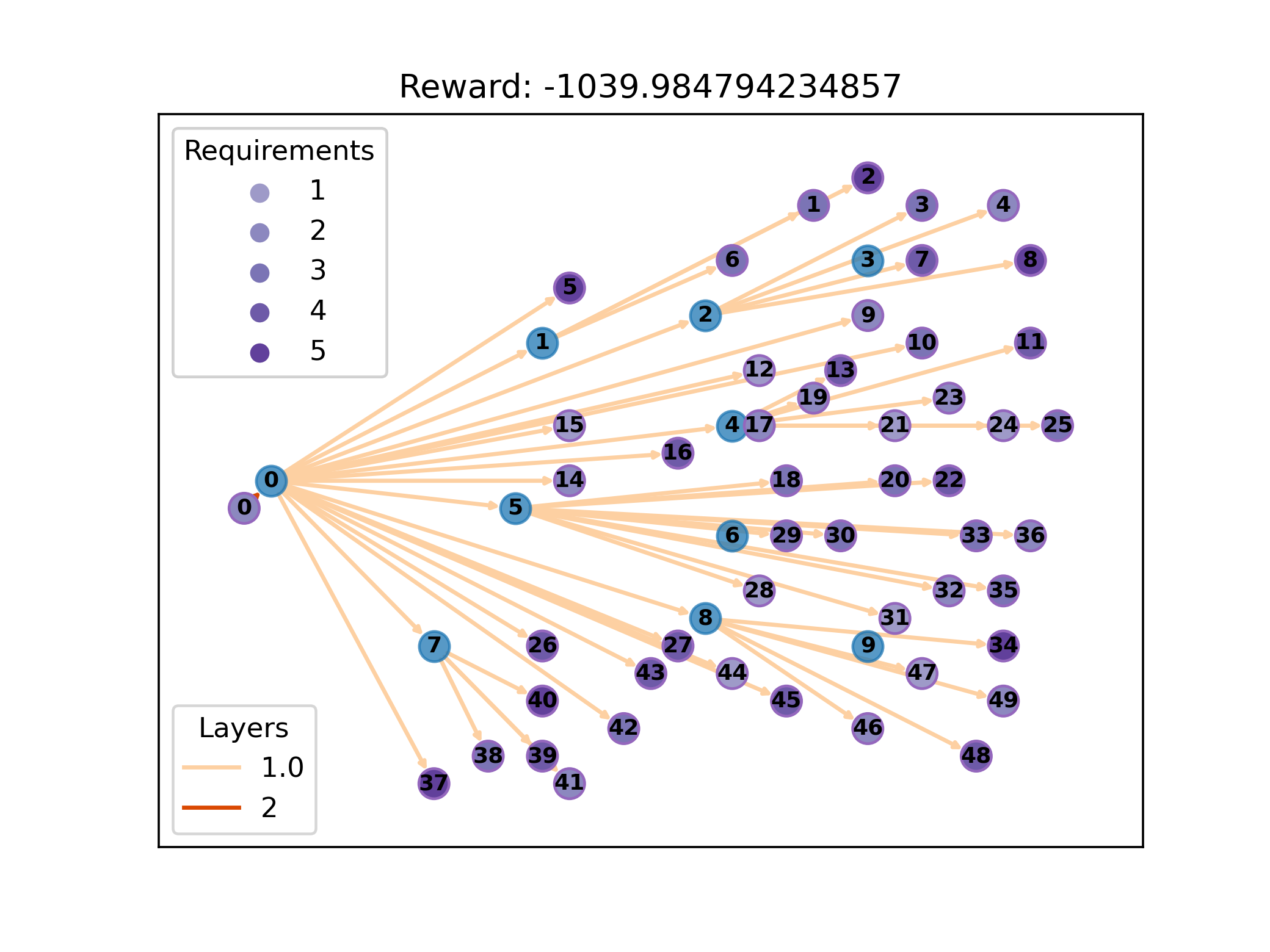}\label{fig:strc_l_qos0}}
~~
  \subfigure[]{\includegraphics[width=0.5\textwidth]{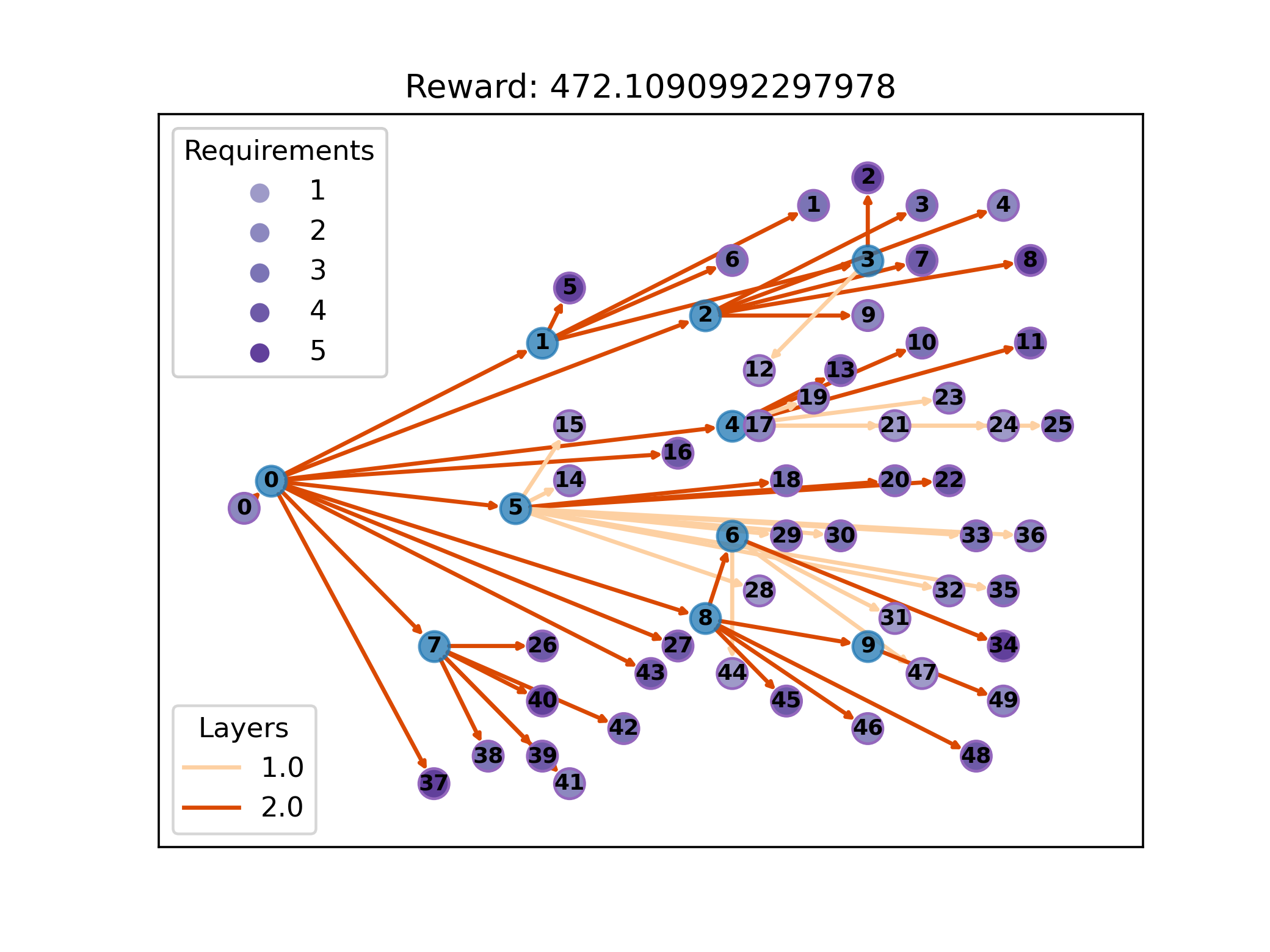}\label{fig:strc_l_qos50}}
~~
  \caption{\name's performance on 10 SFUs, 50 clients structured setting under different QoS parameters. (a) $\alpha=0$. (b) $\alpha=50$.}
  \label{fig:strc_l_qos}
\end{figure*}

Figure \ref{fig:strc_m_qos} shows the different connection topology for the 7 SFUs, 11 clients structured settings under different QoS parameters ($\alpha=0, \alpha=50$). Figure \ref{fig:strc_l_qos} shows the different connection topology for the 10 SFUs, 50 clients structured settings when $\alpha=0$ and $\alpha=50$. 
A higher $\alpha$ means higher preference on bandwidth while a lower $\alpha$ indicates higher preference on latency. 
In both settings \name\ chooses to connect to receiver clients more directly with less intermediate SFUs involved when $\alpha=0$, while more layers are sent when $\alpha=50$, with more SFUs involved in the multicast tree.

\subsubsection{Randomized Settings}
\begin{figure*}[!tbp]
  \centering
  \subfigure[]{\includegraphics[width=0.32\textwidth]{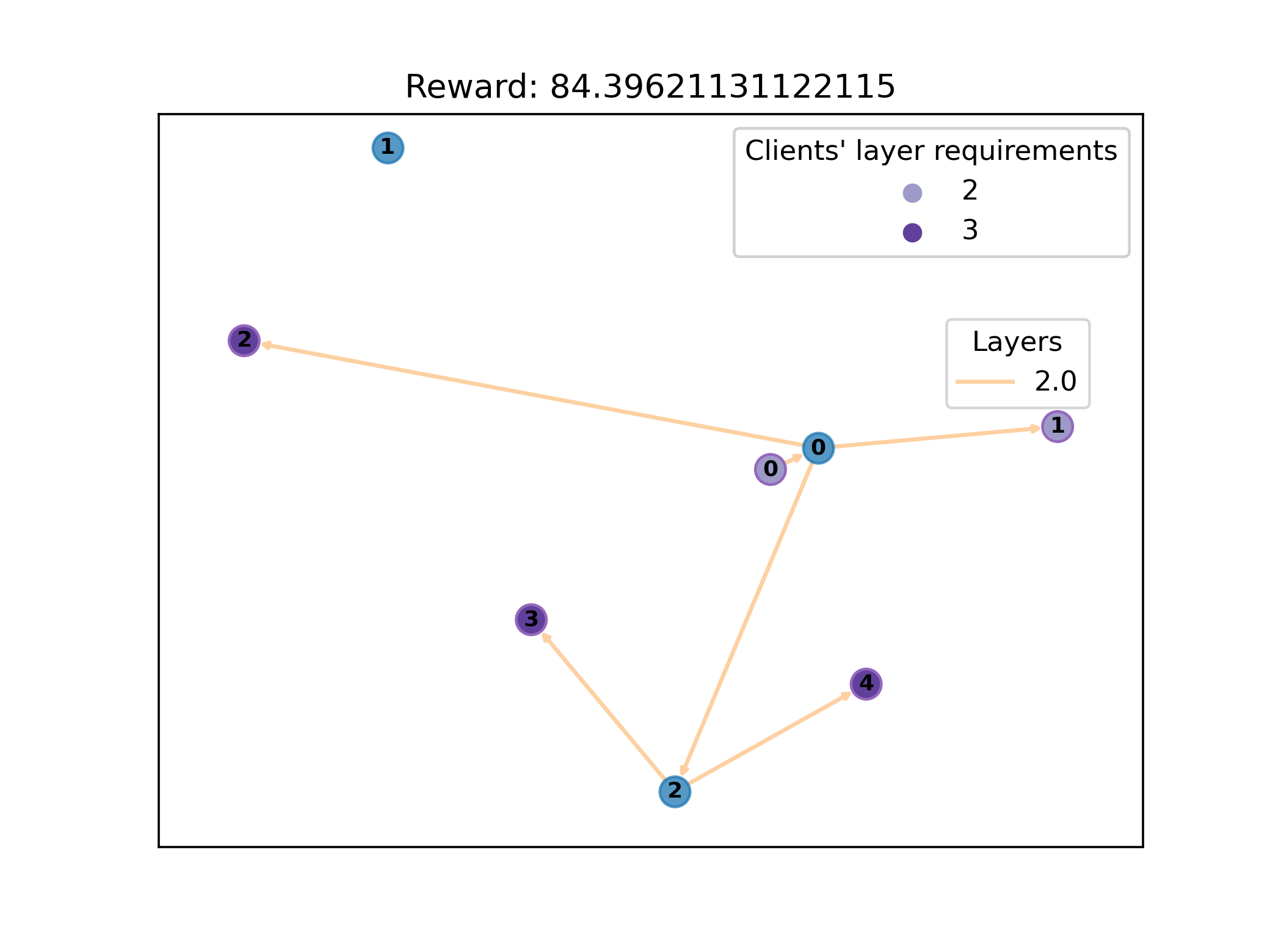}\label{fig:rand_s_1}}
~~ 
  \subfigure[]{\includegraphics[width=0.32\textwidth]{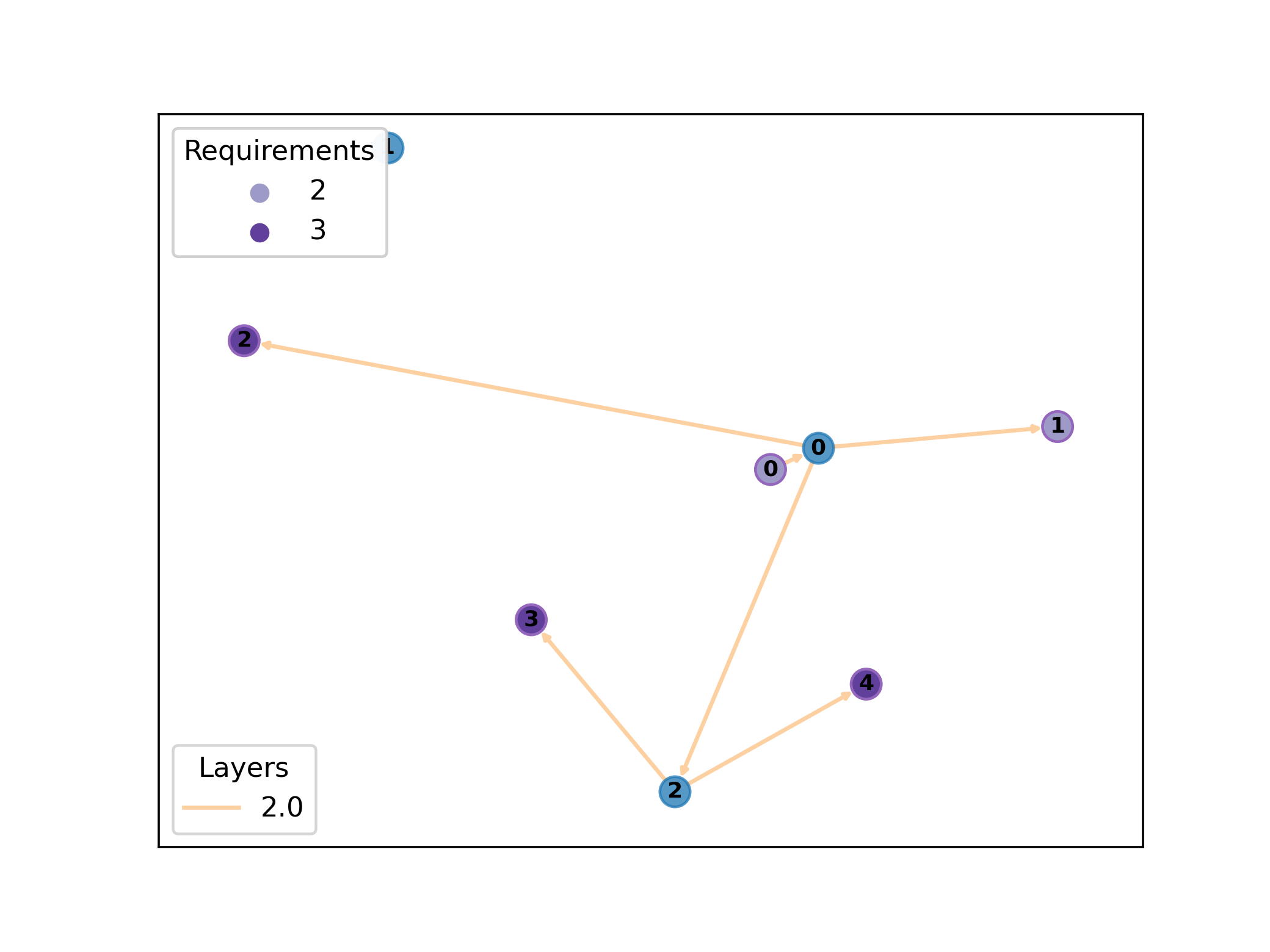}\label{fig:rand_s_2}}
~~  
  \subfigure[]{\includegraphics[width=0.29\textwidth]{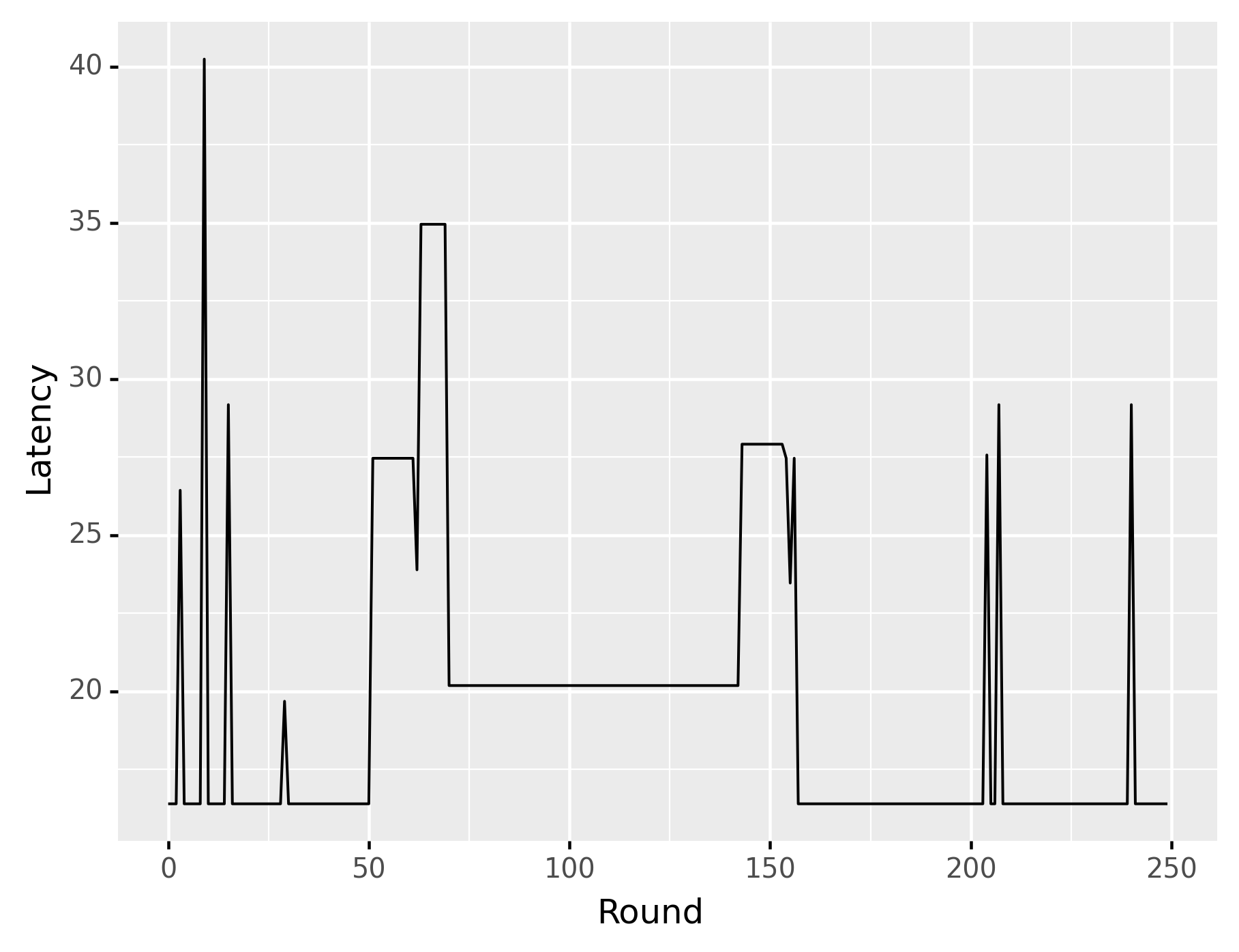}\label{fig:rand_s_3}}
~~
  \caption{A randomly generated setting with 3 SFUs and 5 clients. (a) Connection generated by \name. (b) Connection generated by baseline. (c) Average latency of \name\ with respect to rounds.
  }
  \label{fig:rand_s}
\end{figure*}

\begin{figure*}[!tbp]
  \centering
  \subfigure[]{\includegraphics[width=0.32\textwidth]{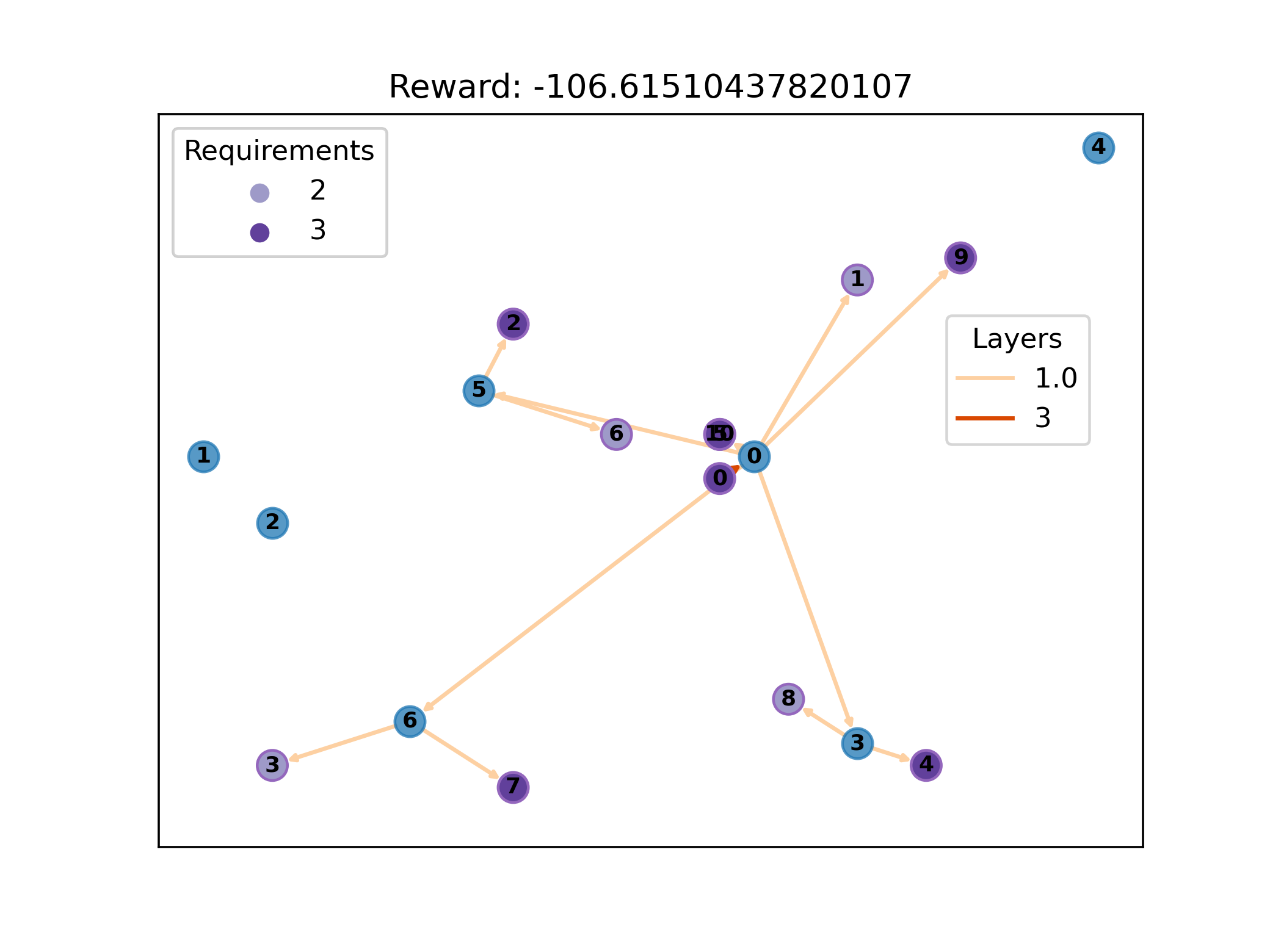}\label{fig:rand_m_1}}
~~ 
  \subfigure[]{\includegraphics[width=0.32\textwidth]{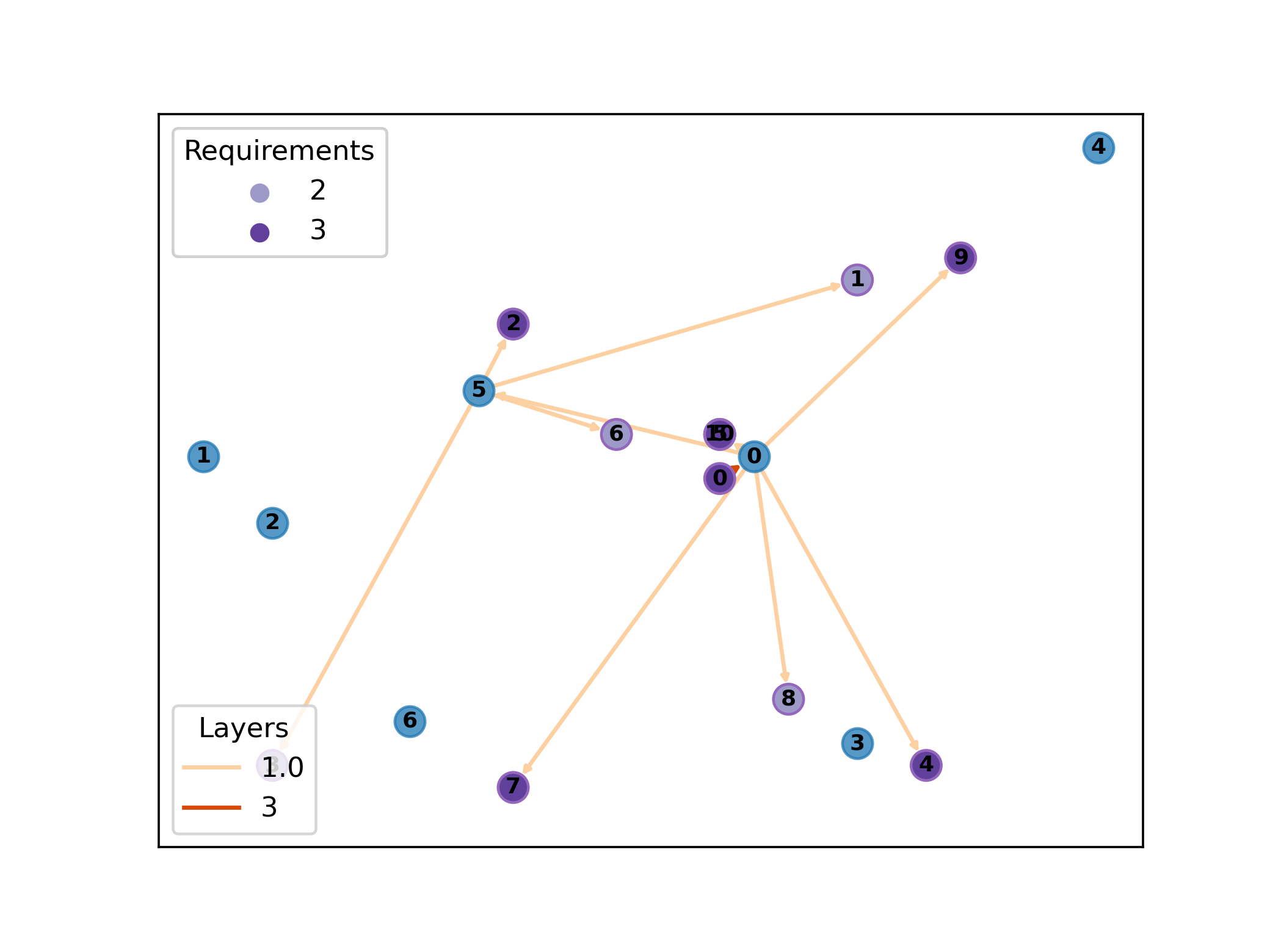}\label{fig:rand_m_2}}
~~
  \subfigure[]{\includegraphics[width=0.29\textwidth]{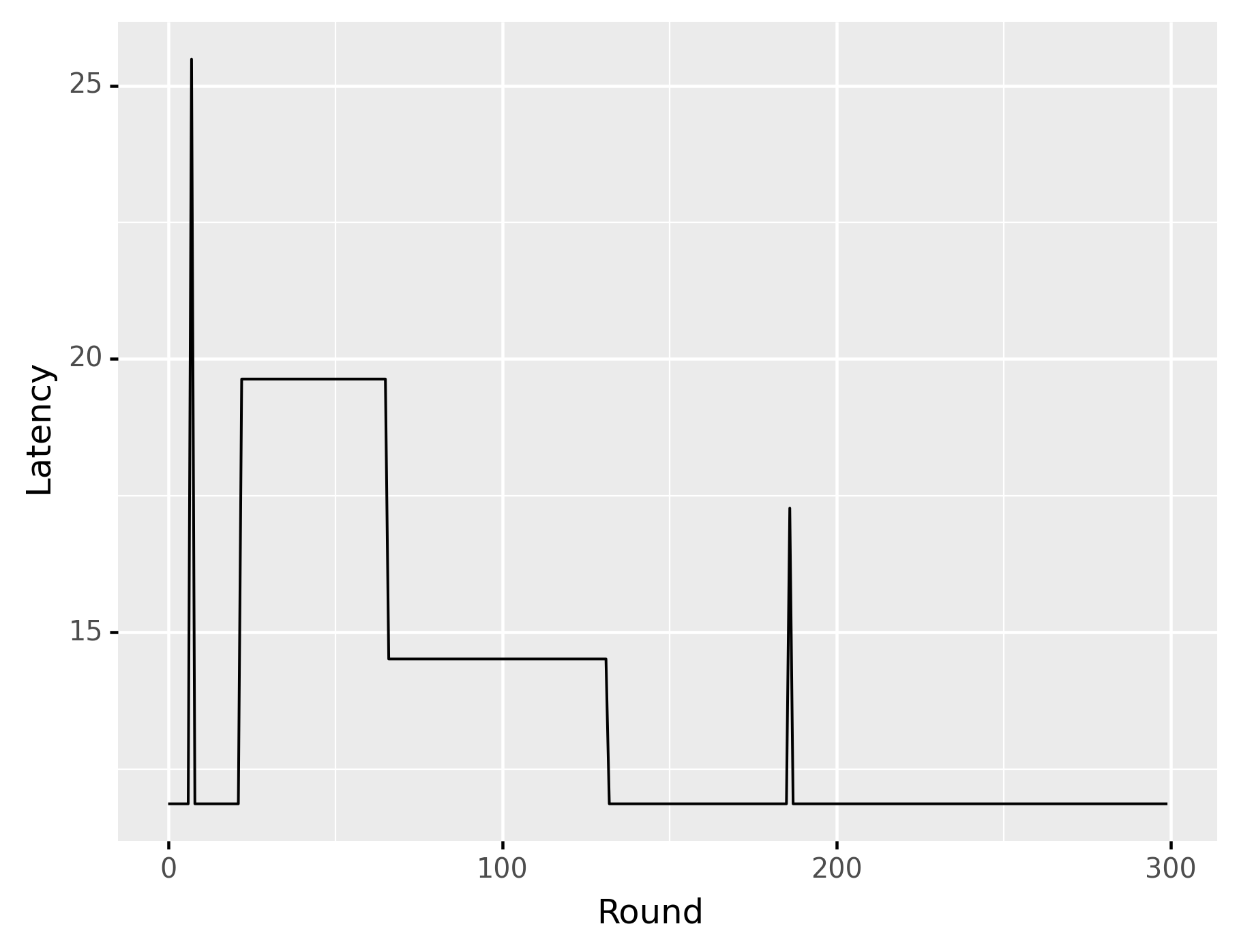}\label{fig:rand_m_3}}
~~
  \caption{A randomly generated setting with 7 SFUs and 11 clients. (a) Connection generated by \name. (b) Connection generated by baseline. (c) Average latency of \name\ with respect to rounds.
  }
  \label{fig:rand_m}
\end{figure*}

\begin{figure*}[!tbp]
  \centering
  \subfigure[]{\includegraphics[width=0.32\textwidth]{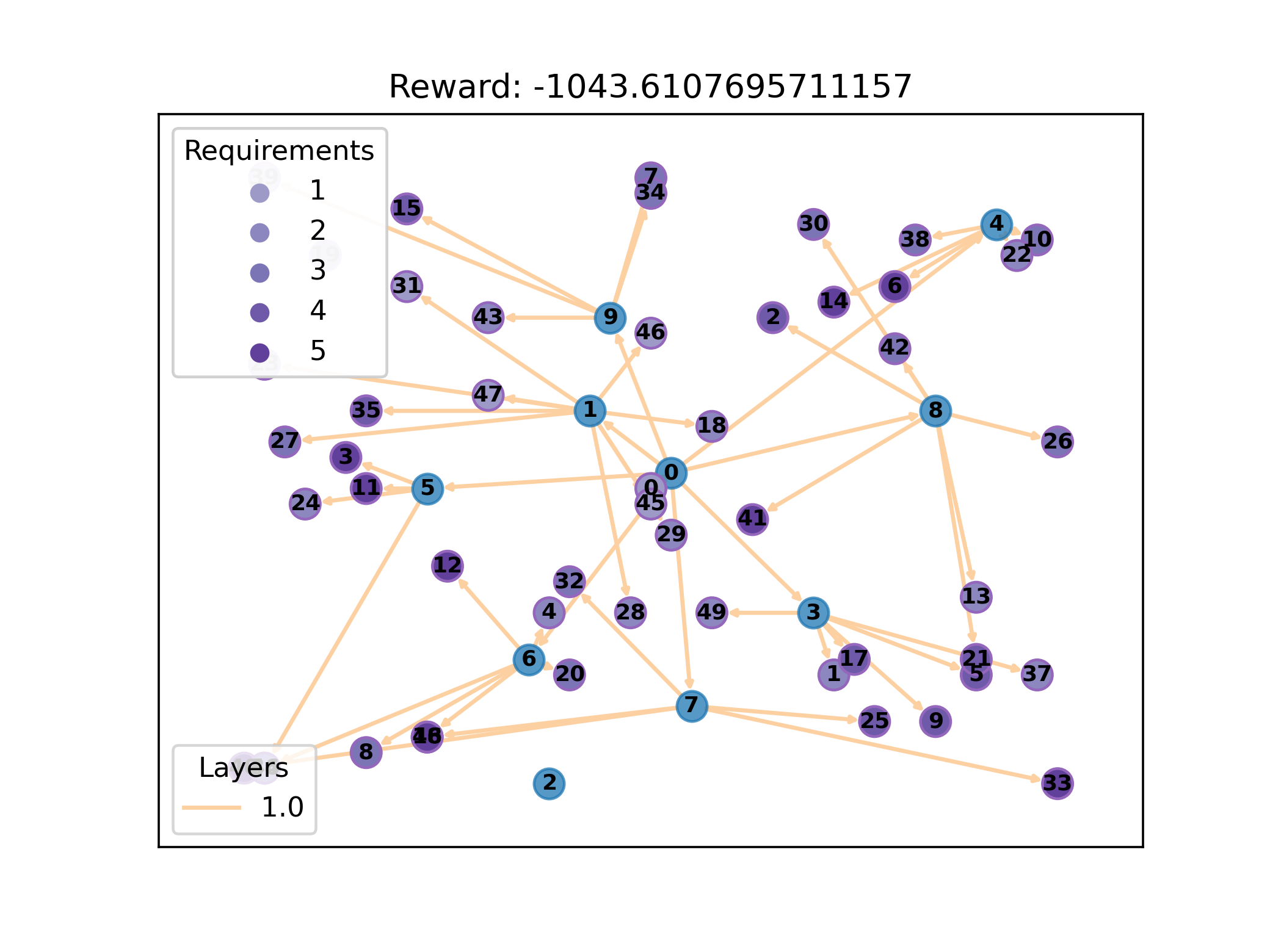}\label{fig:rand_l_1}}
~~
  \subfigure[]{\includegraphics[width=0.32\textwidth]{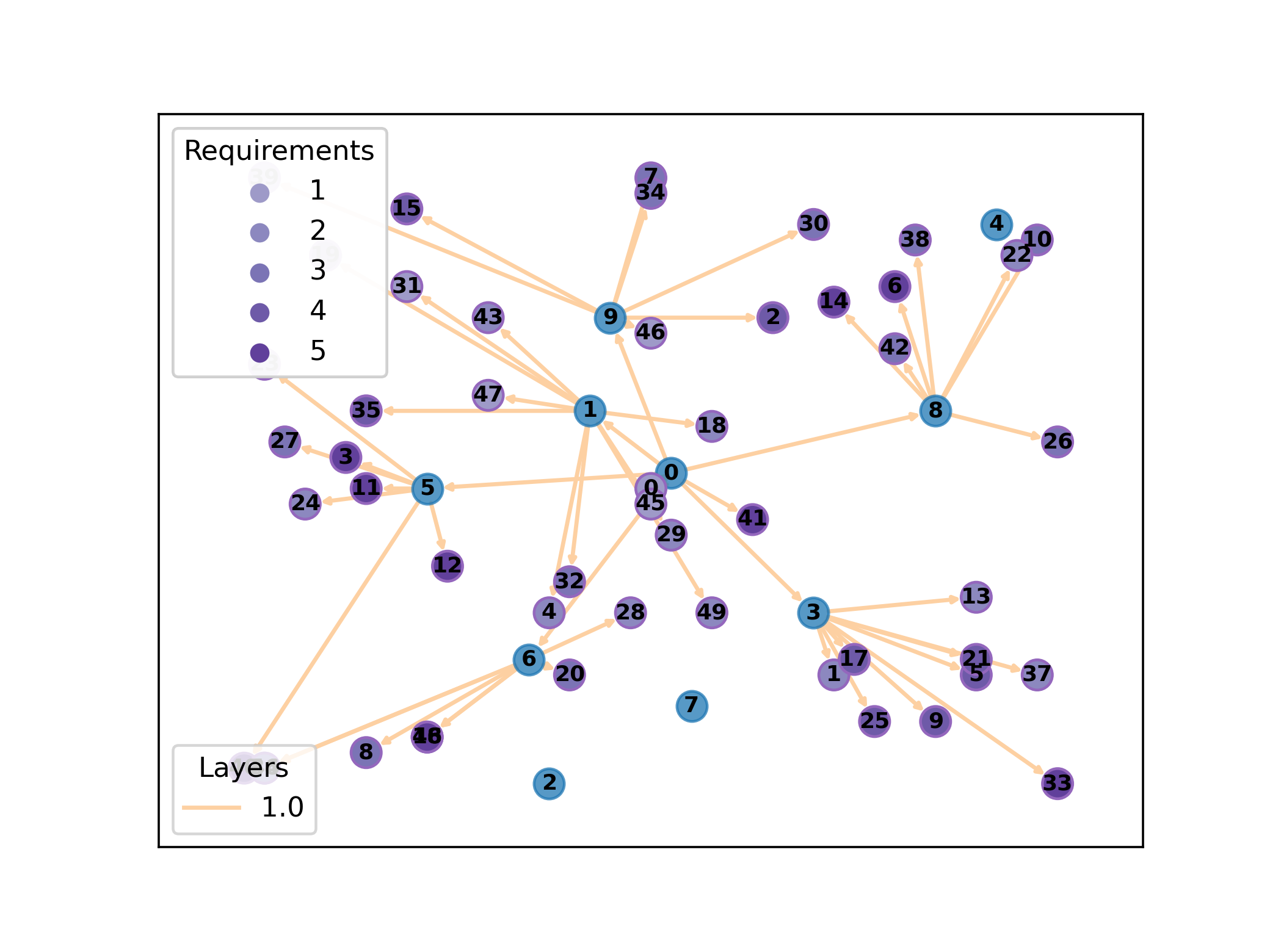}\label{fig:rand_l_2}}
~~
  \subfigure[]{\includegraphics[width=0.29\textwidth]{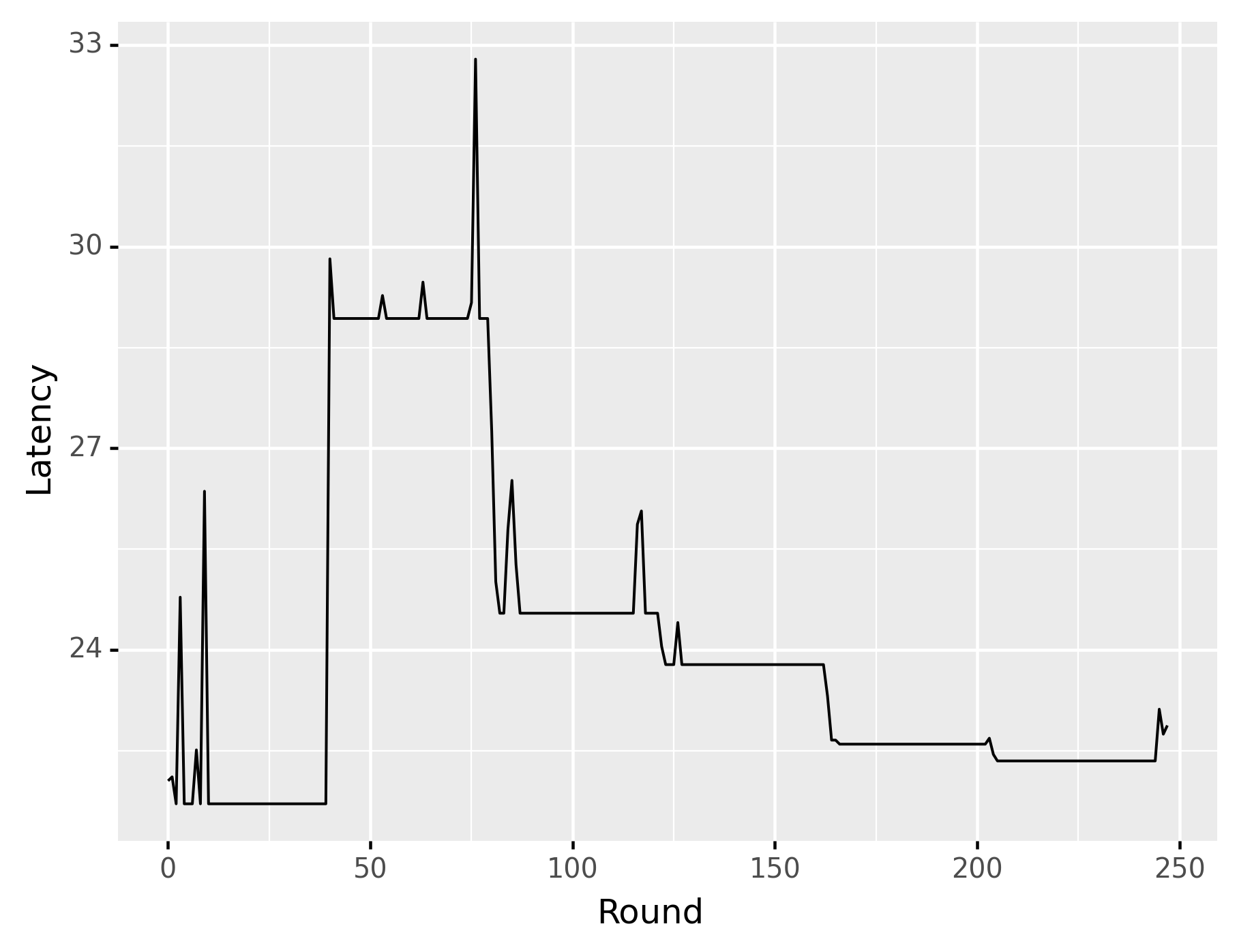}\label{fig:rand_l_3}}
~~
  \caption{A randomly generated setting with 10 SFUs and 50 clients. (a) Connection generated by \name. (b) Connection generated by baseline. (c) Average latency of \name\ with respect to rounds.
  }
  \label{fig:rand_l}
\end{figure*}

Results of randomly generated settings for small, medium and large scales are shown in Figure \ref{fig:rand_s}, \ref{fig:rand_m}, \ref{fig:rand_l}. \name\ achieves the global optimal in the small and medium settings, and gives a connection close to the global optimal in the large setting. Average latency of each receiver clients are shown in (c). The model converges in 200 rounds in the small and medium settings, while it takes about 500 rounds to converge in the large setting.

\subsubsection{A Setting with Multiple Sources Streaming}
\begin{figure*}[!tbp]
  \centering
  \subfigure[]{\includegraphics[width=0.5\textwidth]{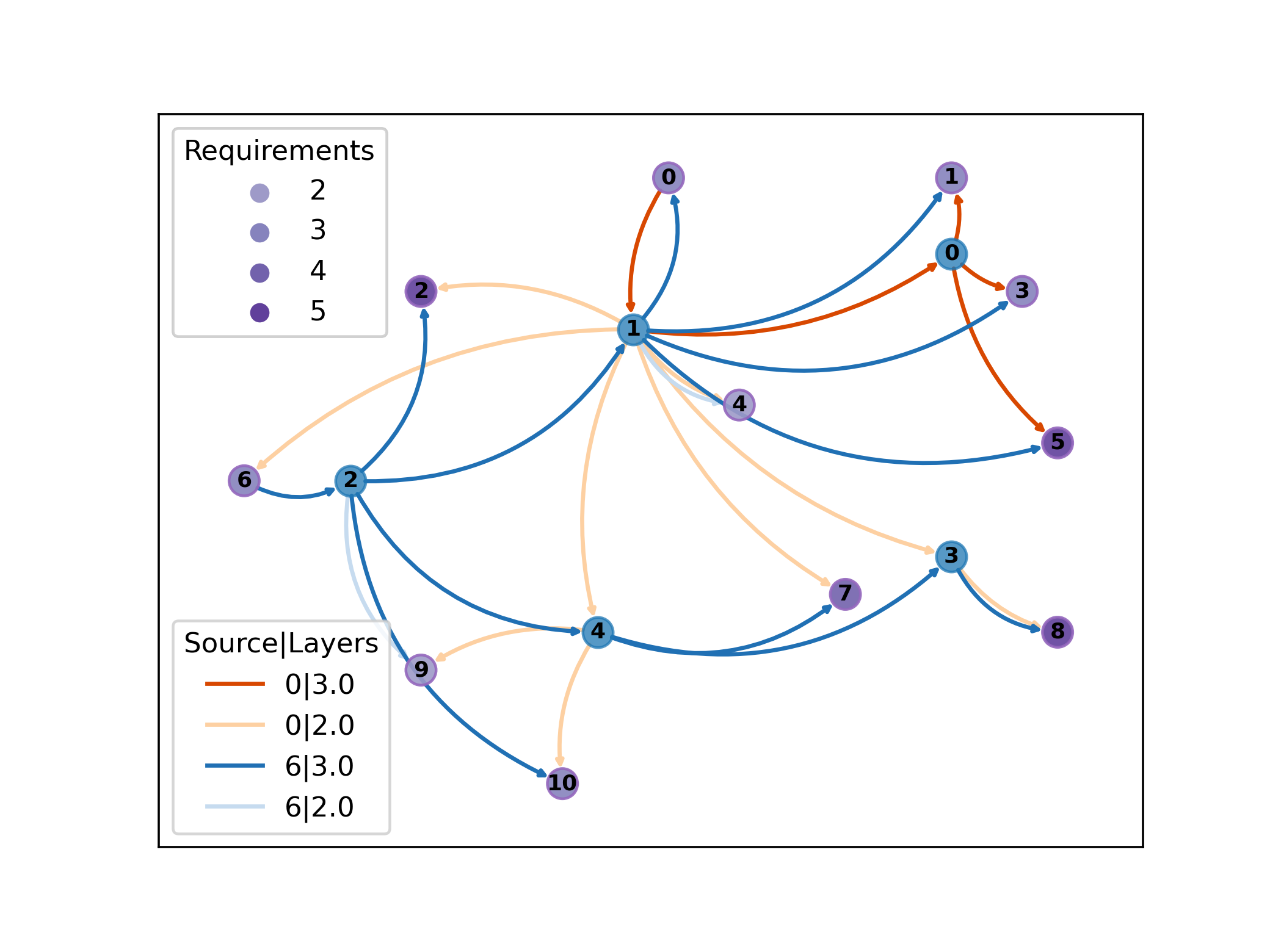}\label{fig:multi_1}}
~~
  \subfigure[]{\includegraphics[width=0.5\textwidth]{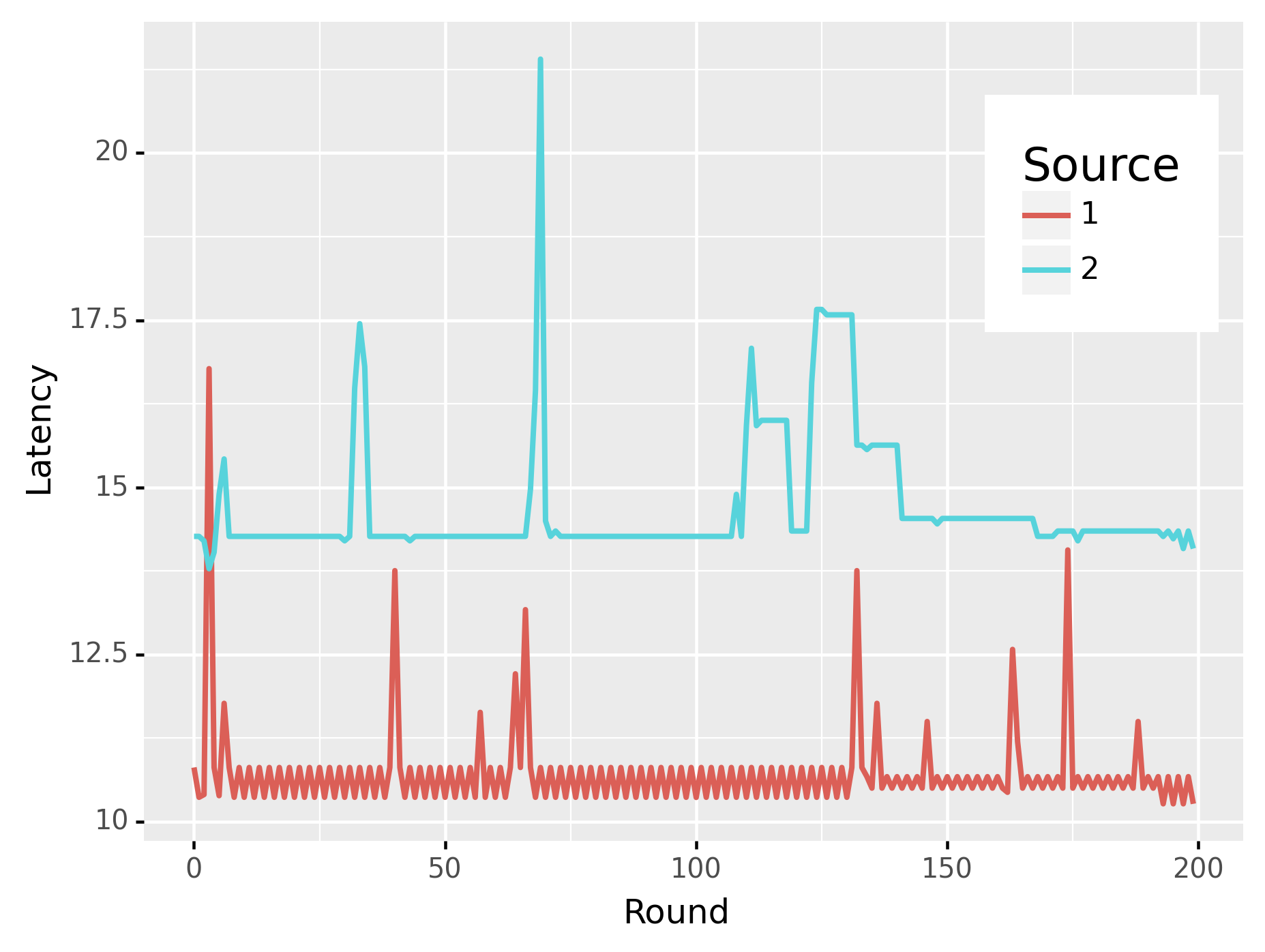}\label{fig:multi_2}}
  \caption{The multiple source streaming setting. (a) Connections generated by \name. (b) Latency with respect to rounds, the source is identified by gateway SFUs, which are SFU 1 and SFU 2.
  }
  \label{fig:multi}
\end{figure*}

To simulate a scenario where multiple users are talking at the same time, a setting is generated (Figure \ref{fig:multi}) with multiple sources sending streams. Two clients (client 0, client 6), are sending streams simultaneously, while all other clients being the receivers (including themselves). The color of the edges in Figure \ref{fig:multi_1} represents the streams sent from different source clients, the shade of color represents the number of layers sent. The performance of latency by each source client with respect to rounds is shown in \ref{fig:multi_2}, given by the average latency from each source client to receiver clients. 

\begin{table}[htbp]
    \centering
    \caption{Location of each SFU and clients.}
    \begin{tabular}{|c|c|c|}
        \hline
        \textbf{Location} & \textbf{Role} & \textbf{Region}\\
        \hline
        Frankfurt & SFU/Data Center & Europe\\
        \hline
        Hong Kong & SFU & Asia\\
        \hline
        San Jose & SFU & North America\\
        \hline
        Paris & Client & Europe\\
        \hline
        Singapore & Client & Asia\\
        \hline
        Taipei & Client & Asia\\
        \hline
        Seattle & Client & North America\\
        \hline
    \end{tabular}
    \label{tab:role_table}
\end{table}

\begin{table*}[]
    \centering
    \caption{Latency by \name\ and common conferencing application model baseline.}
	\begin{tabular}{|c|c|c|c|}
		\hline
		\textbf{Source's Location} & \textbf{Receiver's Location} & \textbf{Baseline's Latency} & \textbf{\name's Latency} \\
		\hline
		\multirow{3}*{Paris} & Seattle & 154.65ms & 169.03ms \\
		\cline{2-4}
		~ & Singapore & 177.64ms & 242.81ms \\
		\cline{2-4}
		~ & Taipei & 248.97ms & 233.29ms \\
		\hline 
		\multirow{3}*{Seattle} & Paris & 154.64ms & 158.24ms \\
		\cline{2-4}
		~ & Singapore & 312.20ms & 183.86ms \\
		\cline{2-4}
		~ & Taipei & 383.53ms & 174.34ms \\
		\hline
	\end{tabular}
	
    \label{tab:latency_table}
\end{table*}

\subsubsection{Real World Video Conferencing Setting}
A real-world video conferencing setting is shown in Table \ref{tab:role_table}. The clients and data centers are located in the cities listed in the table. Clients and data centers in this setting are geographically scattered on different continents so that a scheduling is needed to achieve a good QoE. Latency between every two nodes are acquired from global ping data \cite{globalpingdata}. The proposed model is compared against a centralized baseline model which is the model structure of currently widely used conferencing applications. For this baseline model, it is assumed to involve one data center which serves as the central node: each client sends stream to this node and receives stream from this node. Without the loss of generality, the client located in Paris is chosen to be the host of conference. The central data center is chosen to be the data center geographically closest to the host from the available data centers of the application, which is located in Frankfurt. For \name, we assume that for each client, there exists SFUs (located in San Jose, Frankfurt and Hong Kong) geographically close to it. Two clients, located in Seattle and Paris, are streaming at the same time, with all other participants being the receivers. 

Table \ref{tab:latency_table} shows the latency for each client to receive the streams from the two sources. The average latency for the baseline model is 238.61ms, while for \name\ it is 193.60ms. Compared with the baseline model, \name\ has a lower overall latency since it utilizes the decentralized SFUs world-wide located to help forward the streams.


\section{Related Work}

{\bf P2P video conferencing.}
A number of early works have proposed decentralized algorithms for p2p video conferencing~\cite{liang2011optimal,zhao2013enabling,ponec2009multi}. 
Kurdoglu et al.~\cite{kurdoglu2015dealing} consider a user bandwidth heterogeneity aware peer-to-peer video conferencing model, and design an algorithm to determine video rates in the multi-cast trees. 
The algorithm is, however, centralized while the tree paths have at most two hops.
Kirmizioglu et al.~\cite{kirmizioglu2019multi} also propose a centralized algorithm for multicast tree generation in a setting where video service providers cooperate with network service providers. 
Grozev et al.~\cite{grozev2018considerations} discuss the problem of selecting video conferencing servers based on geographic location. 
This work mainly considers latency in server selection, and is not bandwidth aware.
Another work~\cite{grozev2015last} introduces a stream forwarding strategy in which only streams of the last $N$ active  speakers are forwarded to participants, thus saving bandwidth and CPU processing cost.  

\noindent 
{\bf Combinatorial multi-armed bandits.}
\name\ is inspired by a recent line of  work in combinatorial multi-armed bandits.  
In Chen et al.~\cite{chen2013combinatorial}, the authors propose a combinatorial upper confidence bound (CUCB) algorithm for a setting with stochastic arm rewards assuming existence of an $\alpha, \beta$ approximation oracle to compute the best action each round. 
Qin et al.~\cite{qin2014contextual} present a contextual combinatorial UCB algorithm assuming the reward for each arm is a linear function of its features.  
Chen et al.~\cite{chen2016combinatorial} study a setting where the reward for a super arm depends not only on the mean score of each arm but on the entire distribution.  They propose a stochastic dominant confidence bound algorithm following the optimism under uncertainty principle.  
The combinatorial bandit problem has also been extensively applied to various resource allocation and utility maximization problems. 
E.g., Gai et al.~\cite{gai2010learning} propose algorithms for deciding channel allocations in multiuser cognitive radio networks. 

\noindent 
{\bf Multi-agent bandits and learning.}
Though our work involves multiple agents making bandit decisions, it only has weak connections with existing literature on multi-agent multi-armed bandits. 
Most works on multi-agent bandits assume each agent is working on the same instance of a bandit problem, which is not true in our setting.  
E.g., the works~\cite{madhushani2020dynamic,landgren2021distributed,vial2021robust} consider agents cooperating to find the best arm through communication over a network. 
Wu et al.~\cite{wu2021multi} use a multi-agent bandit algorithm for deciding job allocation in edge computing, but they use the cloud as a centralized coordinator. 
Lauer et al.~\cite{lauer2000algorithm} propose an algorithm for distributed reinforcement learning with independent learners, where each agent has no information about other agent’s behavior, in a cooperative setting where the reward functions are identical for the agents. 
Our algorithm, however, is not cooperative and has combinatorial actions. 
There are few works on the multi-agent combinatorial bandit problem.

\section{Conclusion}
We have presented \name\ a decentralized algorithm for efficient multicast tree construction in open, p2p video conferencing systems. 
Due to the trustless model in open, p2p systems, \name\ is a non-cooperative algorithm that makes actions purely based on its own past observations of the effects of past actions, without collaboration with media servers. 
The adaptive design followed by \name\ finds routing paths that are automatically tuned to the heterogeneities of media servers and clients in where they are located, their processing and bandwidth capabilities etc. without requiring any explicit manual input.   
In practice, additional information about servers may be known such as their reputation score (e.g., derived based on the amount of stake placed on them by other peers~\cite{livepeer}). 
Such information can essentially serve to form prior probabilities for selecting SFUs, and accelerate convergence. 
In many cases, clients have scheduled meetings on a regular or semi-regular basis. 
If a meeting occurs repetitively, information learned from previous meetings may be used to bootstrap the multicast trees.
We leave an evaluation of these techniques for accelerating convergence to future work. 
While servers are not trustworthy in an open network, it may be reasonable to assume clients within a conference session trust each other. 
Designing efficient algorithms where clients mutually share useful information  (e.g., a client can provide a list of SFUs close to her, computed based on ping times to the SFUs, to other clients) is another direction for future work. 
Such interactions between clients can be used to construct a single, global multicast tree (carrying bidirectional traffic), instead of a separate tree per client source, which reduces complexity. 
Our work is motivated by the combinatorial multi-armed bandit problem with time-varying contexts~\cite{zeng2016online}. 
A theoretical understanding of \name's regret in this setting is also an important direction for future work.  

\bibliographystyle{ACM-Reference-Format}
\bibliography{paper}

\appendix

\section{Integer Program Baseline}
\label{s:ipbaseline}
For any nodes $i,j$ with $i \in S \cup \{c_0 \}$ and $j \in S \cup C \backslash c_0$, let $y^{i, j} \in \{0, 1\}$ be a binary variable that denotes whether node $i$ forwards a stream to node $j$, 
$x^{i, j}$ be a non-negative integer denoting the number of layers sent from node $i$ to node $j$, and 
$l(i,j)$ be the latency of sending a packet from $i$ to $j$. 
For $s \in S$, $b_s \in \mathbb{N}$ is the bandwidth limit of SFU $s$. 
For a client $c \in C$, $q^*(c)$ is the number of layers requested by $c$ while $q(c)$ is the number of layers received by $c$. 
For $i \in S, j \in S \cup C, j \neq c_0$ and $c\in C$, $z^{i, j}_c \in \{0, 1\}$ is a binary variable that denotes whether node $i$ makes node $j$ responsible for client $c$. 
$d(c)$ is the latency of the path from $c_0$ to $c$ on the multicast tree, for $c \in C, c\neq c_0$. 
$Q$ is the number of layers the stream is encoded in to at the source $c_0$. 
An integer program to compute an optimal tree can be written as
\begin{align}
    \max &\sum_{\substack{c \in C \\ c \neq c_0}} \left( -d(c) + \alpha \frac{q(c)}{q^*(c)} \right) \\
    \text{such that } \quad &\sum_{\substack{i \in S \\ i \neq j}} x^{i, j} \geq x^{j, k} \quad \forall j \in S, k \in S \cup C, k \neq c_0, j \neq k  \label{const:inoutlayers} \\
    &x^{i, j} \leq Q*y^{i, j} \quad \forall i\in S, j \in S \cup C, j \neq c_0, i \neq j  \label{const:xandyrelation} \\
    &x^{i, j} \geq y^{i, j} \quad \forall i \in S, j \in S \cup C, j \neq c_0, i \neq j \label{const:xandyrelation2} \\
    &\sum_{\substack{j \in S \cup C \\ j \neq c_0, j \neq i}} x^{i, j} \leq b_j \quad \forall i \in S \cup \{c_0\} \label{const:bandwidthx} \\
    &\sum_{\substack{i \in S \cup \{c_0\} \\ i \neq j}} y^{i, j} \leq 1 \quad \forall j \in S \cup C, j \neq c_0 \label{const:oneincomeedge} \\
    &\sum_{i\in S} z^{i, c}_c = 1 \quad \forall c \in C, c\neq c_0 \label{const:zforc} \\
    &\sum_{\substack{i \in S \\ i \neq c_0, i \neq j}} z^{i, j}_c = \sum_{\substack{k \in S \cup C \\ k \neq c_0, k \neq j}} z^{j, k}_c \quad \forall j \in S, c \in C, c\neq c_0 \label{const:zconserv} \\
    &\sum_{\substack{c \in C \\ c \neq c_0}} z^{i, j}_c \geq y^{i, j} \quad \forall i \in S \cup \{c_0\}, j \in S \cup C, j \neq c_0, i \neq j \label{const:zndy} \\
    &\sum_{\substack{j \in S \cup C \\ j \neq c_0, j \neq i}} z^{i, j}_c \leq 1 \quad \forall i \in S \cup \{c_0\}, c \in C, c \neq c_0 \label{const:onlyoneresp} \\
    &\sum_{\substack{c' \in C \\ c' \neq c_0}} z^{c_0, i}_{c'} = (|C|-1)z^{c_0, i}_c  \quad \forall i \in S, c \in C, c \neq c_0 \label{const:c0z} \\
    &\sum_{i \in S} z^{c_0, i}_c = 1 \quad \forall c \in C, c\neq c_0 \label{const:coz2} \\
    &z^{i,j}_c \leq y^{i,j} \quad \forall i \in S \cup \{c_0\}, j \in S \cup C, j \neq c_0, c \in C, c \neq c_0, i \neq j \label{const:zytie} \\
    &q(c) = \sum_{i \in S} x^{i, c} \quad \forall c \in C, c \neq c_0  \label{const:numlayerc} \\
    &q(c) \leq q^*(c) \quad \forall c \in C, c\neq c_0 \label{const:maxlayerreq} \\
    &d(c) = \sum_{\substack{i,j \in S \cup C \\ i \neq j}} z^{i, j}_c l(i, j) \quad \forall c \in C, c \neq c_0 \label{const:disttoc} \\
    &\sum_{\substack{i,j \in T \\ i \neq j}} x^{i, j} \leq |T| - 1 \quad \forall T \subseteq S \cup C, |T| \geq 2 \label{const:tree} \\ 
    & y^{i,j} \in \{0, 1\}, x^{i, j} \in \mathbb{N} \quad \forall i, j \in S \cup C, i \neq j \\
    &z^{i,j}_c \in \{0, 1\} \quad \forall i \in S \cup \{ c_0\}, j \in S \cup C, j \neq c_0, j \neq i, c \in C, c \neq c_0 \\ 
    &d(c) \geq 0, q(c) \in \mathbb{N} \quad \forall c \in C, c \neq c_0 
\end{align}
Eq.~\eqref{const:inoutlayers} says the total number of layers forwarded by an SFU $j$ cannot exceed the number of layers received by $j$. 
Eq.~\eqref{const:xandyrelation} says that if $y^{i, j}$ is zero then $x^{i,j}$ must also be zero, i.e., if $i$ is not connected with $j$ then $i$ must not send any positive number of layers to $j$.  
Similarly Eq.~\eqref{const:xandyrelation2} says that if $y^{i, j} = 1$, then $x^{i, j}$ must also be at least 1. 
Eq.~\eqref{const:bandwidthx} is a bandwidth constraint that requires the total number of layers sent by an SFU to be at most its bandwidth limit. 
Since we are interested in constructing a directed tree, where each node in the tree has exactly one parent node and each node outside of the tree has no parent nodes, we have the constraint in Eq.~\eqref{const:oneincomeedge}. 
Eq.~\eqref{const:zforc} stipulates each client must receive a stream from exactly one SFU. 
An SFU is allowed to make a downstream node reponsible for client $c$ iff the SFU is itself responsible for $c$. 
This requirement is encoded in Eq.~\eqref{const:zconserv}. 
A node $i$ forwads a stream to node $j$ only if node $j$ is responsible for one or more clients, as in Eq.~\eqref{const:zndy}. 
Eq.~\eqref{const:onlyoneresp} says that a node can make at most one downstream node responsible for that client. 
Equations~\eqref{const:c0z} and~\eqref{const:coz2} say that the source $c_0$ forwards the stream to exactly one SFU making that SFU responsible for all clients.  
If node $i$ makes node $j$ responsible for a client, then there must exist a connection between $i$ and $j$. 
This is captured through Eq.~\eqref{const:zytie}. 
The number of layers $q(c)$ received by client $c$ is given by Eq.~\eqref{const:numlayerc}, which must not exceed the number of layers $q^*(c)$ requested by $c$ (Eq.~\eqref{const:maxlayerreq}). 
The overall latency from the source $c_0$ to a client $c$ is given by Eq.~\eqref{const:disttoc}. 
Lastly Eq.~\eqref{const:tree} specifies that the computed routing paths must be a tree (i.e., have no cycles). 
\end{document}